\documentclass[a4paper, 11pt]{article}
\usepackage[utf8]{inputenc}
\usepackage{jcappub}
\usepackage{amsfonts}
\usepackage{graphicx}
\usepackage{amssymb}
\usepackage{amsmath}
\usepackage{breqn}
\usepackage{array}
\usepackage{float}
\usepackage{tikz}
\usepackage{subfig}
\usetikzlibrary{shapes.geometric, arrows}
\tikzstyle{st} = [rectangle, rounded corners, text width = 3cm, text centered, draw = black ]
\tikzstyle{arrow} = [->,>=stealth]

\newcommand{\Mpl}{M_{_{\rm Pl}}}

\def\d{{\rm d}}
\def\x{{\rm \bf x}}
\def\k{{\rm \bf k}}

\def\l{\left}
\def\r{\right}

\title{Inflationary magnetogenesis: solving the strong coupling and its non-Gaussian signatures}

\author{Debottam Nandi}
\affiliation{Department of Physical Sciences, Indian Institute of Science Education and Research (IISER) Mohali, Sector 81 SAS Nagar, Manauli PO 140306, Punjab India.}

\emailAdd{debottam.nandi@gmail.com}

\abstract{
The simplest model of primordial magnetogenesis can provide scale-invariant magnetic fields that can explain the present abundances of it in the cosmic scales. Two kinds of solutions of the coupling function can lead to such phenomena and both of them suffer from the problems of either strong-coupling or large backreaction. In this work, we consider the coupling function as a linear combination of both kinds with a model parameter. We find that the parameter needs to be as small as $\sim 10^{-20}$ in order to evade the backreaction problem. On the other hand, requiring that the modes above Mpc scales do not suffer strong coupling, we also obtain a weak constraint of the model parameter to be greater than $10^{-60}$. For the allowed range of the model parameter, we, then, analytically evaluate the cross-correlation functions between the magnetic fields and the curvature perturbation. We find that such a combination preserves the consistency relation. Also, the result leads to enhanced non-Gaussianity in equilateral as well as flattened limits with unique signatures that characterize the novelty of this model.}

\begin{document}

\maketitle

\section{Introduction}

Astrophysical observations suggest the abundances of large scale coherent magnetic fields in all cosmic scales. These fields are not only present in bound cosmological structures e.g. stars, galaxies, and the cluster of galaxies but they also seem to be present in the intergalactic medium and large scale structures.  In galaxies and clusters of galaxies, the strengths of the magnetic fields have been measured to be a few micro Gauss, they are estimated to be of the order of $10^{-17}$ Gauss in Mpc length scales in intergalactic medium \cite{Neronov_2010, Tevecchio2010, Dermer_2011, Vovk_2012, Tavecchio_2011, Dolag_2010, Takahashi_2011, Huan_2011, Finke_2015} and large scale structures \cite{Grasso_2001, Widrow_2002, Kandus_2011, Widrow_2011, Durrer_2013,Ade:2015cva, Ade:2015cao}.

The origin of such magnetic fields is still unknown.  Various explanations that have been put forward can broadly be classified into two categories: Astrophysical and Primordial \cite{Grasso:2000wj, Widrow:2002ud, Subramanian:2009fu, Subramanian:2015lua}. The primary issue with the astrophysical sources is that, they cannot explain the coherent length of the field. It is therefore believed that the magnetic field is
generated in the early universe and are then amplified by the galactic dynamo mechanism at late times to give rise to such strength of the fields in those scales.

Inflationary paradigm \cite{STAROBINSKY198099, Sato:1981, Guth:1981, LINDE1982389, Albrecht-Steinhardt:1982, Linde:1983gd, Mukhanov:1981xt, HAWKING1982295, STAROBINSKY1982175, Guth:1982, VILENKIN1983527, Bardeen:1983, Starobinsky:1979ty} is currently is the most successful early Universe paradigm that can explain the Cosmic Microwave Background Radiation (CMBR) and the formation of large scale structures. However, during this time, the standard electromagnetic (EM) field decays very fast due to its conformal invariance structure in the four dimensions and thus it can still not generate sufficient seed magnetic fields. Therefore, it has been realized that to generate such strength, the necessary condition is to break the conformal symmetry of the EM field. Several authors have suggested many ways to break the conformal invariance of the EM field by introducing (i) a coupling of the EM field with the Ricci/Riemann tensors, (ii) a non-minimal coupling of the EM field with scalar/axion/fermionic field, and (iii) compactification from higher dimensional space-time \cite{Ratra:1991bn, 1990-Kolb.Turner-Book, Dolgov:1993vg, Gasperini:1995dh, Davidson:1996rw, Calzetta:1997ku, Prokopec:2001nc, Giovannini:2000dj, Bamba:2003av, Bamba:2006ga, Demozzi:2009fu, Campanelli:2008kh, Kanno:2009ei, Subramanian:2009fu, Urban:2011bu, Durrer:2010mq, Byrnes:2011aa, Jain:2012jy, Kahniashvili:2012vt, Ng:2015ewp, Bamba:2014vda, Tasinato:2014fia,Fujita:2015iga, Campanelli:2015jfa, Fujita:2016qab, Kamada:2020bmb, Bamba:2020qdj, Maity:2021qps, Sharma:2017eps, Sharma:2018kgs}. Recently, a new model has also been introduced where the conformal symmetry is broken due to the higher derivative action of the $U(1)$ field without any coupling function that can generate sufficient magnetic field \cite{Nandi:2017ajk}.

In this work, we are interested in models where the standard EM action is coupled with the scalar field responsible for the slow-roll inflation. The time-dependent coupling function leads to the amplification of EM field fluctuations and thus can generate an ample amount of magnetic field. The coupling function is chosen in such a way that, at the end of inflation, it becomes unity and later remains the same so that the standard EM theory is restored. There exist two solutions of the coupling function that lead to the scale-invariant magnetic field which is preferred. However, even with the success, these kinds of models suffer two crucial problems: the strong coupling problem and the backreaction problem \cite{Demozzi:2009fu, Ferreira:2013sqa, Ferreira:2014hma}. The backreaction problem arises due to the fact that the energy density of the EM field grows faster than the background energy density, thus can break the inflationary dynamics, which is not preferred. The second is the strong coupling problem in which the time evolution of the coupling function during inflation leads to the EM field being in the strong coupling regime at the beginning of inflation. This implies that the perturbative calculation of the EM field fluctuations can not be trusted.

In this work, we try to construct the coupling function in such a manner that the scale-invariant spectrum of the magnetic field remains the same, and at the same time, the above problems can also be evaded. Since the solutions are dual, the obvious choice is to choose the coupling function as a linear combination of both solutions. This kind of choice provides two parameters of which, one can be fixed by using the fact that, coupling function becomes unity at the end of inflation. This leads to one free model parameter. By choosing such a combination, one can immediately see that the inverse of the coupling function, i.e., the \emph{effective coupling constant}, while going backward in time from the end of inflation, grows to a maximum, and at then decays very fast, unlike the conventional case, and thus it \emph{may} address the issue of the strong coupling. Also, by demanding that there is no backreaction at the largest observable scales, we may immediately obtain constraints on this parameter.

In order to search for its implications, it would be interesting to study the cross-correlations between the magnetic fields and the curvature perturbation. This has been extensively studied in the literature \cite{Caldwell:2011ra, Motta:2012rn, Jain:2012ga, Jain:2012vm, Kunze:2013hy, Chowdhury:2018blx, Chowdhury:2018mhj}. It is found that the magnitude of non-Gaussianities generated through such correlations has been estimated to be quite large for inflation. Also, it is found that the consistency relation: whether the non-Gaussian parameter in the squeezed limit can be written in terms of the two-point function, is satisfied \cite{Jain:2012ga, Jain:2012vm}. Motivated by these results, we also study the same and found that our new choice of the coupling function preserves the consistency relation, irrespective of the choice of the model parameter. Also, the non-Gaussianity parameters in the other limits are obtained, which are different than the conventional scenario studied before. The results indicate an interesting outlook of such choice of the viable coupling function.

The article is organized as follows. In the following section, we discuss the basic model of the primordial magnetogenesis. We obtain the preferred scale-invariant solution of the magnetic field that can explain $\sim 10^{-12}$ G strength of the field in Mpc scales at present time. We also discuss the two main problems associated with this kind of solution. In Sec. \ref{sec:proposedmodel}, we introduce the new coupling function and obtain the energy density of the EM field. Requiring no backreaction in the furthest observable scales, we obtain the constraint on the model parameter. We also show that the observable modes, i.e., the modes associated with the Mpc length scales exits the Hubble horizon and become \emph{classical} before the system enters into the strong coupling regime, thus evading the problem for the modes we are interested in. In Sec. \ref{sec:crosscorr}, we then briefly discuss the cross-correlations of the magnetic fields and the curvature perturbation and evaluate the non-Gaussianities associated with the new coupling function in Sec. \ref{sec:ng}. We show that, while the consistency relation is preserved, it is highly enhanced in other limits with unique signatures. The results also differ from previous works and we conclude our work with a brief discussion in Sec. \ref{sec:conclu}.

A few words on our conventions and notations are in order at this stage of our discussion. In this work, we work with the natural units such that $\hbar=c=1$, and we define the Planck mass to be $\Mpl =(8\,\pi\, G)^{-1/2}$. We adopt the metric signature of $(-, +, +, +)$. Also, we should mention that, while the Greek indices are contracted with the metric tensor $g_{\mu \nu}$, the Latin indices contracted with the Kronecker delta $\delta_{i j}$. Moreover, we shall denote the partial and the covariant derivatives as 
$\partial$ and $\nabla$. The overdots and overprimes, as usual, denote derivatives with respect to the cosmic time~$t$ and the conformal time~$\eta$ associated with the Friedmann-Lema\^{\i}tre-Robertson-Walker (FLRW) line-element, respectively.

\section{The simple model of primordial magnetogenesis}\label{sec:simplemdel}

In this section, we briefly discuss the mechanism to generate magnetic fields in the early Universe. Let us first consider the background to be the spatially flat, Friedmann-Lema\^itre-Robertson-Walker (FLRW) metric that is described by the line-element
\begin{equation}\label{eq:FLRW metric}
\d s^2 = a^2(\eta)\, 
\l(-\d\eta^2+\delta_{ij}\, \d x^i\,\d x^j\r),
\end{equation}
where $a(\eta)$ is the scale factor and $\eta$ denotes the conformal time coordinate, and $\delta_{i j}$ is the Kronecker delta function. The standard electromagnetic theory is given by the action:
\begin{equation}\label{eq:standard EM action}
S_{\rm EM} = -\frac{1}{4}\int {\rm d}^{4}x\, \sqrt{-g}\, F_{\mu\nu}F^{\mu\nu}, \quad F_{\mu \nu} \equiv \partial_\mu\,A_\nu - \partial_\nu\,A_\mu,
\end{equation}

\noindent where $A_{\mu}$ is the vector potential and the above action is gauge-invariant. The electric and the magnetic fields are defined as

\begin{equation}\label{eq:define E B}
	E_{i}=-\frac{1}{a}A_{i}^{\prime} , \qquad B_{i}=\frac{1}{a}\epsilon_{ijk}\partial_{j}A_{k}
\end{equation}
The action is conformally invariant in four dimensions. As a result, one can show that the EM field fluctuations decay as $a^{-2}$ with the expansion of the universe, and therefore cannot explain the present abundances of the magnetic field. Hence, one needs to break the conformal invariance for the amplification of the vacuum fluctuations. There exist infinite possibilities, yet the simplest and the most elegant one is to couple the action \eqref{eq:standard EM action} with a field responsible for the early Universe dynamics, that can also preserve the gauge invariance:
\begin{equation}\label{eq:primordial EM action}
S = -\frac{1}{4}\int \d^{4}x\, \sqrt{-g}\, f^{2}\left(\phi\right)F_{\mu\nu}F^{\mu\nu},
\end{equation}
where $\phi$ is a homogeneous scalar field which is assumed to be the inflaton field. $f(\phi)$ is the coupling function responsible to break the conformal symmetry. Since gauge invariance is preserved, one can choose an arbitrary gauge. In this work, we use the Coulomb gauge where $A_{0}=0$ and $\partial_{i}A^{i}=0$ and the action \eqref{eq:primordial EM action} becomes
\begin{equation}\label{eq:primordial EM coulomb}
S=\frac{1}{2}\int \d^{3}\x\, \d\eta\, f^{2}\left(\phi \right)\left(A_{i}^{\prime}{}^{ 2}-\frac{1}{2}\left(\partial_{i}A_{j}-\partial_{j}A_{i}\right)^{2}\right).
\end{equation}

\noindent The above action leads to the equation of motion of the vector field $A_i$ as

\begin{equation}\label{eq:EOM vector}
A_{i}^{\prime \prime} + 2\,\frac{f^\prime}{f} A_{i}^\prime - \partial_{j}\partial^{j}A_{i}=0.
\end{equation}

\noindent In order to quantize the EM field, the vector potential can be Fourier decomposed as
\begin{equation}\label{eq:Fourier expansion}
\hat{A}_{i}\left(\eta, \x \right)=\sum_{\sigma=1,2}\int\frac{\d^{3} \k}{\left(2\pi\right)^{3/2}}\,\epsilon_{i\,\sigma}^{\k}\,\left(\hat{b}^\sigma_{\k}\,A_{\k}\left(\eta\right)\,e^{i\,\k.\x}+\hat{b}^{\sigma \dagger}_{\k}\,A^\ast_{\k}\left(\eta\right)\,e^{-i\,\k.\x}\right),
\end{equation}

\noindent where $\epsilon_{i\,\sigma}^{\k}$ is the polarization vector and $\sigma$ corresponds to the spin associated with it. $\hat{b}^\sigma_{\k}$ and $\hat{b}^{\sigma \dagger}_{\k}$ are the annihilation and creation operator. By definition $\delta_{ij}{\epsilon}^{i}{\epsilon}^{j}=1$
and the following identities are verified: 

\begin{equation}
	{\epsilon}_{\sigma}^{i}k_{i}=0, \quad \sum_{\sigma=1,2}{\epsilon}_{\sigma}^{i}\left(\mathbf{k}\right){\epsilon}_{j,\sigma}\left(\mathbf{k}\right)=\delta_{j}^{i}- \delta_{jl}k^{i}k^{l}/{k^{2}},\nonumber
\end{equation}
with the usual commutation relation
\begin{equation}\label{eq:commutation}
\left[\hat{b}_{\mathbf{k_{1}}}^{\sigma},\hat{b}_{\mathbf{k_{2}}}^{\sigma'\dagger}\right]=\delta^{\left(3\right)}\left(\mathbf{k_{1}}-\mathbf{k_{2}}\right) \delta_{\sigma \sigma'}.
\end{equation}

\noindent Let us now define the canonically normalized vector field ${\cal A}_{k} \equiv f(\eta)\,A_{k}(\eta)$ (dropping the vector sign as the field depends only on the amplitude of $\k$). Rewriting the action \eqref{eq:EOM vector} in terms of the canonical field gives us
\begin{equation}\label{eq:EOMcanonicalVec}
{\cal A}_k^{\prime\prime}+\left(k^{2}-\frac{f^{\prime\prime}}{f}\right){\cal A}_{k}=0.
\end{equation}

\subsection{Energy spectrum}

In this section, we focus on the energy spectrum, which later will help up to evaluate the amount of the magnetic field generated during the early Universe. It will also be crucial to understand the backreaction problem of such kinds of models. Varying the action \eqref{eq:primordial EM action} with respect to the metric provides us the energy-momentum tensor of the EM field:

\begin{equation}\label{eq:energymomentum}
T_{\mu\nu}=f^{2}(\phi)\,\left(F_{\mu}^{\ \beta}F_{\nu\beta}-\frac{1}{4}g_{\mu\nu}F_{\alpha\beta}F^{\alpha\beta}\right).
\end{equation}

\noindent The above energy-momentum tensor leads to the total energy density of the EM field as

\begin{equation}\label{eq:total enrgy-density}
	\rho_{\rm EM}=\rho_{\rm E}+\rho_{\rm B},
\end{equation}
where $\rho_{E}$ and $\rho_{B}$ are the energy densities of the Electric and magnetic fields, respectively:
\begin{eqnarray}\label{eq:ElecDen}
	\rho_{\rm E} &\equiv& \frac{f^2(\phi)}{2} g^{i j} E_i E_j = \frac{f^2(\phi)}{2\,a^4(\eta)} A_i^\prime{}^2, \\\label{eq:MagDen}
	\rho_{\rm B} &\equiv& \frac{f^2(\phi)}{2} g^{i j} B_i B_j = \frac{f^2(\phi)}{4\,a^4(\eta)} \left(\partial_{i}A_j - \partial_{j}A_i\right)^2.
\end{eqnarray}

\noindent Using \eqref{eq:Fourier expansion} and the above expressions, one can immediately obtain the electric and magnetic field spectra as

\begin{eqnarray}
	\mathcal{P}_{\rm B} \equiv \frac{d\rho_{B}}{d\log k} &=& \frac{1}{2\pi^{2}}\frac{k^{5}}{a^{4}}\left|{\cal A}\left(k,\eta\right)\right|^{2}\label{eq:MagSpec},\\
	\mathcal{P}_{\rm E} \equiv\frac{d\rho_{E}}{d\log k}&=&\frac{f^{2}}{2\pi^{2}}\frac{k^{3}}{a^{4}}\left|\left(\frac{{\cal A}\left(k,\eta\right)}{f}\right)^\prime\right|^{2}.\label{eq:ElecSpec}
\end{eqnarray}

\noindent Since, at the end of inflation, the coupling becomes unity, the magnetic field at the end of inflation is simply:

\begin{equation}\label{eq:mag end inf}
	B_{\rm end}(k, a) = \sqrt{\left.\mathcal{P}_{\rm B}(k, a)\right|_{\rm end}},
\end{equation}
where, $|_{\rm end}$ implies the value at the end of inflation.
\subsection{Inflationary solution}

During slow-roll inflation, the scale factor solution can \emph{approximately} be written as a de-Sitter solution:
$a(\eta) \approx -\frac{1}{H \eta}$, where $H$ is the Hubble parameter which is nearly constant. There are infinite possibilities of choosing the coupling function, however, the simplest of them is to choose the power-law solution of the scale factor, i.e., $$f \propto a^\alpha \propto (-\eta)^{-\alpha}.$$ Assuming the slow-roll inflatinary Universe is governed by the canonical scalar field minimally coupled to gravity with potential $V(\phi)$, the coupling function in terms of the scalar field can be obtained as $$f(\phi) \propto e^{-\alpha\int \frac{V \d\phi}{V_\phi}},$$  where $V_\phi \equiv \partial V/\partial \phi.$ After the end of inflation, the Universe lies at the minima of the potential. Therefore, to restore the standard EM theory, the function needs to be unity, which can easily be achieve with appropriate normalization. For example, consider the potential $V(\phi) \propto \phi^2$, i.e., the chaotic potential, and the coupling function for this model becomes $f(\phi) = e^{-\alpha \phi^2/4}$. This corresponds to the approximate time-dependent solution of the coupling function as 
\begin{equation}\label{eq:couplsol}
	f(\eta) = \left\{\begin{aligned}&(\eta/\eta_{\rm end})^{-\alpha}, \quad \eta \leq \eta_{\rm end} \\
	&0, \qquad \qquad \,\,\,\eta > \eta_{\rm end},
	\end{aligned}\right.
\end{equation}
where $\eta_{\rm end}$ is the time when inflation ends. With such choice, Eq. \eqref{eq:EOMcanonicalVec} becomes

\begin{equation}
{\cal A}_{k}''+\left(k^{2}-\frac{\alpha\left(\alpha+1\right)}{\eta^{2}}\right){\cal A}_{k}=0,\label{eq:Simplified equation of motion for a power law coupling}
\end{equation}
whose solution can be written in terms of the Bessel functions as
\begin{equation}\label{eq:solpowervecfield}
{\cal A}_{k}=\sqrt{-k\eta}\,\left[C_1(k) J_{\alpha + 1/2} (- k \eta) + C_2(k) J_{-\alpha - 1/2} (- k \eta)\right].
\end{equation}
By imposing the Bunch-Davies initial conditions, the two integration constants $C_1$ and $C_2$ can easily be fixed as
\begin{equation}
	C_{1}\left(k\right)=\sqrt{\frac{\pi}{4k}}\frac{e^{-i(\alpha-1)\pi/2}}{\cos\left(\alpha\pi\right)}, \qquad
	C_{2}\left(k\right)=\sqrt{\frac{\pi}{4k}}\frac{e^{i\alpha \pi/2}}{\cos\left(\alpha\pi\right)}.
\end{equation}

\noindent When the modes leave the Hubble horizon, i.e., when $k \eta \ll 1$, one can use the asymptotic solution of the Bessel's function and using that, the above solution can be written as

\begin{equation}\label{eq:solasmp}
 \left.{\cal A}_{k}\right|_{-k\eta \rightarrow 0}= \frac{C_1(k)}{2^{\alpha + \frac{1}{2}}\Gamma(\alpha + \frac{3}{2})} (-k \eta)^{1 + \alpha} + \frac{C_2(k)}{2^{-\alpha - \frac{1}{2}}\Gamma(\frac{1}{2} - \alpha)} (-k \eta)^{ - \alpha}.
 \end{equation}

\noindent This implies that, as $k\eta \rightarrow 0$, for $\alpha > -1/2$, the first part of the right-hand side of the solution dominates, whereas the second part dominates for $\alpha < -1/2$. 

\subsection{Scale-invariant Magnetic spectrum and the present abundances}
 
Since we obtain the solutions of the primordial magnetic field, with the help of \eqref{eq:MagSpec}, one can find out the magnetic field energy density at the end of inflation. At the super-Hubble scale, the spectrum \eqref{eq:MagSpec} becomes

\begin{equation}
	\mathcal{P}_{\rm B} = \frac{\mathcal{F}(n)}{2 \pi^2} H^4 (-k \eta)^{4 + 2n}
\end{equation}

\noindent where $n = -\alpha, \quad \alpha > -1/2$ and $n = 1 + \alpha, \quad \alpha < -1/2$, and $\mathcal{F}(n) = \pi/(2^{2n + 1}\Gamma^2(n + 1/2)\cos^2(n \pi))$ \cite{Subramanian:2015lua}. We have also used $a(\eta) \approx - 1/(H\eta)$ as the near de-Sitter solution for the slow-roll inflationary scenario. It is then obvious that the scale-invariant magnetic field can be obtained for $\alpha = 2$ and $-3$ and in these cases, the spectrum $\sim \delta^2_{\rm B}$ becomes $\mathcal{O}(1) H^4$. This implies that the magnetic field generated at the end of inflation is $$\delta_{\rm B} \sim H^2.$$ From the CMB observations \cite{Akrami:2018odb}, one can find that the constraint on the Hubble scale at the of inflation is $H \sim 10^{-6} \Mpl$ which is translated to the strength of the magnetic field to be $10^{-12}$ in Planck scale or $10^{45}$ G at the end of inflation.

After the end of inflation, since the coupling function becomes unity and the standard EM theory is restored, Eq. \eqref{eq:EOMcanonicalVec} loses the term $f^{\prime\prime}/f$. This implies that, after the end of inflation $\delta_{\rm B}$ decays as $a^{-2}$ and the electric field vanishes quickly \cite{Subramanian:2015lua}. Assuming instantaneous reheating and using the entropy conservation, one can obtain how much universe expands after the end of inflation to the present epoch and it is approximately

$$\left(\frac{a_0}{a_{\rm end}}\right) \sim 10^{29},$$  
where $a_0$ is the scale factor at present time. Since the magnetic field decays as $a^{-2}$, the field value today cannot exceed $10^{-13}$ G.

\subsection{Backreaction and strong coupling problem}

Let us now focus on the backreaction problem. In order to ensure that there is no backreaction due to the EM field, the energy density (see Eq. \eqref{eq:total enrgy-density}) must remain smaller than the background energy density $\rho_{\rm inf} \equiv 3 H^2 \Mpl^2$.  The main contribution to the energy density comes from the super-Hubble scales because the contribution from the subhorizon scales is renormalized in the leading order. Then, similar to the magnetic field spectrum, one can simply obtain the electric field spectrum for the super-Hubble modes as

\begin{equation}
	\mathcal{P}_{\rm E} = \frac{\mathcal{G}(m)}{2 \pi^2} H^4 (-k \eta)^{4 + 2m},
\end{equation}
where, $m = 1 - \alpha, \quad \alpha > 1/2$ and $m = \alpha, \quad \alpha < 1/2$, and $\mathcal{G}(m) = \pi/(2^{2m + 3}\Gamma^2(m + 3/2)\cos^2(\pi m)).$ The energy density of the electric field can be obtained as

\begin{eqnarray}
	\rho_{\rm E} &=& \int_{a_i H}^{a_{\rm end} H} \mathcal{P}_{\rm E}(k)\frac{\d k}{k} \nonumber \\
	&=& \left\{\begin{aligned}
	&\frac{\mathcal{G}(m)}{2 \pi^2 (4 + 2 m)} H^4 \left(1  - \left(\frac{a_i}{a_{\rm end}}\right)^{4 + 2m}\right), \quad m \neq -2, \\
	& \frac{\mathcal{G}(m)}{2 \pi^2} H^4\log\left(\frac{a_i}{a_{\rm end}}\right), \quad m = -2,
	\end{aligned}\right.
\end{eqnarray}
where $a_i$ is the scale factor at the beginning of inflation.

Now, consider the case of scale-invariant magnetic field spectrum. This can be obtained for $\alpha = -3$ and $2$. For $\alpha = -3$, the density corresponding to the electric field becomes $$ \mathcal{O}(1) H^4 \left(1  - \left(\frac{a_{\rm end}}{a_{i}}\right)^{2}\right).$$ Assuming inflation lasts at least for 60 e-folds time, $|\rho_{E}|/\rho_{\rm inf} \approx 10^{52} H^2/\Mpl^2 \approx  10^{40} \gg 1$. It implies that the electric field density dominates over the inflaton field energy density $\rho_{\rm inf}$ and the backreaction spoils the homogeneous background. Therefore, the solution corresponding to $\alpha = -3$ is not allowed.  However, one can evaluate the same with $\alpha = 2$ and show that, in that case, the electric field energy density remains ${\cal O}(1)  H^4$ which is much smaller than that of the background field. Therefore, the allowed scale-invariant solution for magnetogenesis is $\alpha = 2$.

However, unlike the curvature perturbations, the magnetic field energy density does not need to be scale-invariant and the sole purpose of magnetogenesis is to generate sufficient magnetic fields. Therefore, one can ask, for what value of $\alpha$, one can get the maximum amplitude of the magnetic field without the backreaction problem. It can easily be shown that the value is $\alpha \simeq 2.2$ for which the magnetic field at the present epoch at Mpc scale can be as high as $10^{-8}$ G \cite{Demozzi:2009fu}.

However, these allowed range of the coupling function $f(\phi)$ possesses another subtle yet crucial problem: strong coupling. Consider the vector field is coupled with a charged fermion in the standard form:

$${\cal L} = -\frac{1}{4} F_{\mu \nu} F^{\mu \nu} + i \bar{\psi}\gamma^\mu (\partial_\mu + i g A_\mu) \psi$$

\noindent where $g$ is the coupling constant, then after re-scaling the vector potential by the coupling
constant $A_\mu \rightarrow g A_\mu,$ we bring this Lagrangian to the form:

$${\cal L} = -\frac{1}{4 g^2} F_{\mu \nu} F^{\mu \nu} + i \bar{\psi}\gamma^\mu (\partial_\mu + i A_\mu) \psi,$$ which indicates that the coupling function $f$ acts like the inverse of the coupling constant. Therefore, small values of $f$ correspond to large coupling constant $g$, which implies that the theory lies in the uncontrollable strong coupling region and the quantum theory cannot be trusted: known as the strong coupling problem. Therefore, only larger values of $f$, i.e., $f \geq 1$ is allowed in order to avoid the strong coupling problem \cite{Demozzi:2009fu}.

Therefore, for $\alpha \geq 2,$ the coupling function at the beginning of inflation takes the value $f_i \simeq 10^{-52} - 10^{-57},$ which indicates that these `allowed' values suffers from strong coupling problem. In fact, it has been shown that one can generate maximum amplitude of the magnetic field for the value of $\alpha \simeq - 2.2$ \emph{without both strong coupling and backreaction problems}, and in that case, the field generated today cannot exceed $10^{-30}$ G in Mpc scales \cite{Demozzi:2009fu}. 

\section{The proposed model: choosing a new coupling function}\label{sec:proposedmodel}

In the above section, we discussed how we can generate magnetic fields with the coupling function that breaks the conformal symmetry of the standard EM theory. As mentioned above, up to $10^{-8}$ G magnetic field amplitude can be generated that can successfully answer the existence of magnetic fields in large scales. However, again, as discussed, it severely suffers from the strong coupling problem and the theory cannot be trusted. 

Consider the scale-invariant case $\alpha = 2$, i.e., $f \propto (-\eta)^{-2}$. While this choice can generate the required primordial field, the value of it gets extremely small toward the beginning of inflation. On the other hand, for $\alpha = 3$, while there is no strong coupling, it spoils the background energy density as the EM field causes backreaction.

However, it should be noted that the solution with strong coupling problem does not possess backreaction and vice-versa, and motivated by this, in this article, we choose the coupling function as the linear combination of the above two scale-invariant cases, i.e.,

\begin{equation}
	f(\eta) = d_1 \left(\frac{\eta}{\eta_{\rm end}}\right)^{3} + d_2 \left(\frac{\eta}{\eta_{\rm end}}\right)^{-2},
\end{equation}
where $d_1, d_2$ are constants. Requiring $f \rightarrow 1$ at the end of inflation helps to re-write the above as

\begin{equation}\label{eq:new choice coupling}
	f(\eta) = d \left(\frac{\eta}{\eta_{\rm end}}\right)^{3} + (1 - d) \left(\frac{\eta}{\eta_{\rm end}}\right)^{-2}.
\end{equation}

\noindent We have re-written $d_1$ as $d$, which is again a constant. The novelty of such choice is mainly because the two solutions are dual, i.e., they both lead to the similar solution of the canonical field ${\cal A}_k$, and therefore, any linear combination of them can also lead to a similar solution, which helps in evaluating everything analytically. Then, one can also ask for its implications in the context of primordial magnetogenesis. In the next section, we will show that in order to solve the backreaction, $d$ must be smaller than $10^{-20}$, which, in turn, can evade the problem of strong coupling for modes corresponding to Mpc scales.

\subsection{Solving backreaction problem}\label{sec:solv_back}

For such choice of the coupling function, Eq. \eqref{eq:new choice coupling} leads to $f^{\prime\prime}/f = 6/\eta^2$, which corresponds solution of the canonical vector field ${\cal A}_k$, i.e., Eq. \eqref{eq:solpowervecfield} for $\alpha = 2$. Note that the solution is identical to the solution $\alpha = -3$ as $f^{\prime\prime}/f$ is identical to that of $\alpha=2$, i.e., the two solutions are dual. In this case, the present day magnetic field strength can be generated up to $10^{-13}$ G, as discussed before. Now, let us focus on the electric field spectrum. In the super Hubble scale, using the expression \eqref{eq:ElecSpec}, it becomes

\begin{equation}
	\left.\mathcal{P}_{\rm E}\right|_{ -k \eta \rightarrow 0} = \frac{H^4}{4 \pi^2}\frac{1}{x^2}\frac{\left(-15\, d\, x^4 + (1- d)\, x\, x_{\rm end} \right)}{d\, x^5 + (1 -d)},
\end{equation}
where $x \equiv -k\eta,~ x_{\rm end} \equiv - k \eta_{\rm end}$. We can now directly evaluate the energy density of the electric field as previously done and it becomes

\begin{equation}
	\rho_{\rm E} \simeq \frac{225\,d^2\,H^4}{8 \pi^2}\left(\frac{a_{\rm end}}{a_i}\right)^2.
\end{equation}
The total energy density of the EM field is dominated by the electric field density as the magnetic field energy density becomes logarithmic for the solution and thus can be ignored. Requiring the total EM energy density must be smaller than the background energy density $3H^2 \Mpl^2$, one can find the constraint on the parameter $d$ as:

\begin{equation}
	d < \frac{2 \sqrt{2}\,\pi}{5\sqrt{3}}\frac{\Mpl}{H}\left(\frac{a_i}{a_{\rm end}}\right)
\end{equation}

\noindent Assuming inflation lasts for at least $60$ e-folds, $d$ must be smaller than $10^{-20}.$

This is one of the main results of this work: \emph{we have considered the linear combination of the dual scale-invariant solution and find the constraint on such combination.} In the next section, we will show the implications of such choice on the strong coupling problem.

\subsection{Solving strong coupling problem}
In order to generate $10^{-13}$ G magnetic field today in Mpc scales, the coupling function becomes smaller and smaller at earlier times, and at the beginning of the inflation, it reaches the value $10^{-52}$, implying that the effective coupling constant is huge. In this section, we show that the above choice of the coupling function \eqref{eq:new choice coupling} can avoid such a high value of the coupling constant and remains very briefly in the strong coupling region. This region also lies close to the end of the inflation, where all observable modes are already far outside the Hubble radius, thus avoid the strong coupling problem.

In the previous section, we show that the magnitude of $d$ needs to be extremely small, i.e., $d \ll 1$ in order to avoid the backreaction. Because of such choice, as one can see from \eqref{eq:new choice coupling}, as $|\eta|$ increases, initially $f$ decays as $\eta^{-2}$ and enters into the strong coupling region. Shortly after that, it reaches a minimum and then it grows as $\eta^3$ and quickly recovers from the strong coupling problem. One can easily estimate the region for which the proposed coupling function remains in the strong coupling region and when it gets out of that region. The minimum can easily be found by equating the time derivative of the coupling function to zero and assuming $d \ll 1$, it becomes 

\begin{equation}
	\left(\frac{\eta_{\rm min}}{\eta_{\rm end}}\right) \sim d^{-1/5}, \qquad f_{\rm min} \sim f(\eta_{\rm min}) \approx d^{2/5}.
\end{equation}

\begin{figure}
	\centering
	\includegraphics[scale=.65]{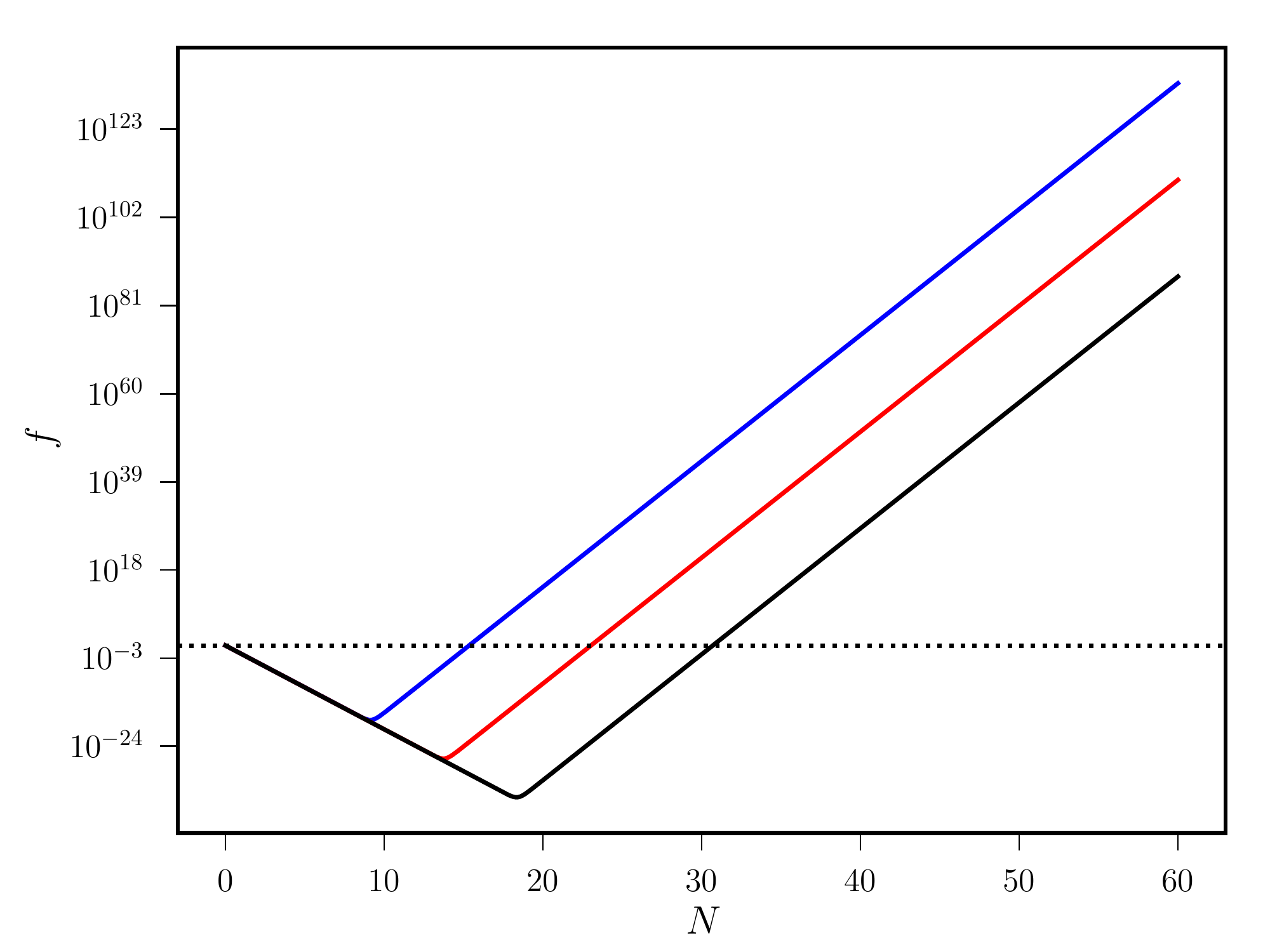}
	\caption{The proposed coupling function is plotted against the inflationary e-folding number $N$. 0 denotes the end of inflation, whereas positive $N$ values denote the earlier past, and blue, red and black curve correspond to $d = 10^{-20},~10^{-30}$ and $10^{-40}$, respectively. From this figure, it is evident that, while in the range $\eta_{\rm end}\leq\eta\leq\eta_{\rm min}$, the coupling function $f$ acts as $\eta^{-2}$, it behaves as $\eta^3$ for $\eta \geq \eta_{\rm min}.$ Also, among these three examples, $d = 10^{-40}$ makes the system stay in the strong coupling region (when $f < 1$) for the longest period $N < 30,$ while the modes corresponding to Mpc scales leave the Hubble horizon $N \sim 50.$ \label{fig:coupling}}
\end{figure}

\noindent $\eta_{\rm min}$ is time when the coupling function reaches the minimum. This implies that, the coupling function behaves as $\eta^{-2}$ in the region $|\eta_{\rm end}| \leq |\eta| \leq d^{-1/5} |\eta_{\rm end}|$ and the minimum value of the coupling function is $d^{2/5}$. To get the perspective, consider $d \approx 10^{-20}$ with the assumption that the CMB pivot, i.e., $0.05$ Mpc scale exits the Hubble radius 50 e-fold before the end of inflation, we get $|\eta_{\rm end}| \sim 10^{-20}$ Mpc, $|\eta_{\rm min}| \sim 10^{-16}$ Mpc and $f^{-1}_{\rm min} \sim 10^8$. However, after $|\eta| > |\eta_{\rm min}|,$ the coupling function starts growing as $\eta^3$ and quickly reaches unity at $$\left(\frac{\eta_\ast}{\eta_{\rm end}}\right) \sim d^{-1/3}.$$ $\eta_\ast$ is the time when the coupling function reaches unity, again. This implies that the model remains in the domain of strong coupling for $|\eta_{\rm end}|\leq |\eta| < d^{-1/3}|\eta_{\rm end}|$ with the peak of the coupling constant $f^{-1} \sim d^{-2/5}$. For similar choice of $d$ to be $10^{-20}$, we get the domain of strong coupling as $10^{-20}\, {\rm Mpc} \le |\eta| \le 10^{-13}$ Mpc.  Note that the region gets broader with the choice of smaller values of $d$ with higher coupling constant $f^{-1}_{\rm min}.$ Therefore, it is always preferred to choose the $d$ as large as possible. This becomes apparent from Fig. \ref{fig:coupling}. 

Compared to the conventional scenario with $f \propto \eta^{-2},$ the above example thus reduces the problem of strong coupling drastically, as the domain of the strong coupling and the amplitude of the coupling constant $f^{-1}_{\rm min}$ gets smaller than before. The above example also shows that, compared to the Mpc scales, $\eta_\ast$ lies very close to the end of inflation. At this time, the modes corresponding to Mpc scales become super horizon and \emph{classical}, similar to the primordial perturbations. Requiring the modes corresponding to the Mpc length scales leave the Hubble radius before the model enters into the region of strong coupling \emph{weakly} provides the lower bound of $d$ as $\sim 10^{-60}.$ Since the strong coupling problem relates to whether the quantum origin of the EM field can be trusted, one can argue that, for $10^{-60} < d < 10^{-20},$ the problem does not arise as the observable modes become classical, thus such choice of coupling function \emph{solves the strong coupling problem for modes of Mpc scale.}

This is the main result of this work: \emph{we introduce a new coupling function, which corresponds to the linear combination of the dual scale-invariant magnetic field solution and find that it can evade the strong coupling problem for modes of Mpc scales,  which we are interested in, and simultaneously can produce $10^{-13}$ G magnetic field in Mpc scales today without any backreaction.} In the next section, we will study the effect of such model in higher-order cross-correlation functions.

\section{Cross-correlations between magnetic fields and curvature perturbation}\label{sec:crosscorr}

In the previous section, we showed that the proposed coupling function can evade the problem of strong coupling for Mpc scales, and at the same time, can provide sufficient seed magnetic field. However, it is evident that, from the study of power spectrum, we are unable to detect new observational information, and therefore, the obvious thing to do is to study the higher-order cross-correlation functions between the magnetic fields and the curvature perturbation. This has been extensively studied in the literature \cite{Caldwell:2011ra, Motta:2012rn, Jain:2012ga, Jain:2012vm, Kunze:2013hy, Chowdhury:2018blx, Chowdhury:2018mhj}. In this section, we extend the calculation for our model.

Let us start with the quadratic action for the curvature perturbation $\mathcal{R}$ that provides the evolution of the mode functions. The action is

\begin{equation}
	\mathcal{S}_{\mathcal{R}} = \int \d^3\x\, \d \eta  \,a^2(\eta)\,\epsilon \,\left(\mathcal{R}^\prime{}^2 - \partial_{i}\mathcal{R}\partial^{i}\mathcal{R}\right),
\end{equation}

\noindent where, $\epsilon \equiv -\frac{\dot{H}}{H^2}$ is the slow-roll parameter. We can write the curvature perturbation $\mathcal{R}$ in terms of the mode functions as

\begin{equation}
	\mathcal{R}(\eta, \x) = \int \frac{\d^3 \x}{(2 \pi)^{3/2}} \left(\hat{a}_{\k} \mathcal{R}_\k(\eta) e^{i \k.\x} + \hat{a}^\dagger_{\k} \mathcal{R}^\ast_\k(\eta) e^{-i \k.\x}\right)
\end{equation}
and during slow-roll inflation, the solution of the mode function in leading order slow-roll can be written as

\begin{equation}\label{eq:mode_curvature}
	\mathcal{R}_k(\eta) = \frac{1}{\sqrt{2 \,\epsilon}}\frac{H}{\sqrt{2 \,k^3}} \left(1 + ik\eta\right)e^{-ik\eta}.
\end{equation}

\noindent Now, in order to evaluate the cross-correlations between magnetic fields and curvature perturbation, we first need the interaction Hamiltonian involving the vector fields $A_i$ and curvature perturbation $\mathcal{R}$. This can easily be obtained from the third order perturbed Lagrangian as the density of the interaction Hamiltonian $\mathcal{H}_{\rm int} = -\mathcal{L}_3$ (see Refs. \cite{Nandi:2015ogk, Nandi:2016pfr, Nandi:2017pfw}):

\begin{equation}\label{eq:int_ham}
	H_{\rm int} = \frac{1}{2}\int \d^3\x \,f^2(\eta) \left[\left(\mathcal{R} - \frac{\mathcal{R}^\prime}{a H}\right)\left(A_i^\prime{}^2 + F_{i j} F^{i j} \right) + \frac{2\partial_{i}\mathcal{R}}{a H}A_j^\prime F^{i j}\right].
\end{equation}

\noindent The cross-correlation between the magnetic fields and the curvature perturbation in real space is defined as

\begin{eqnarray}
	\left<\mathcal{R}(\eta, \x) B_i(\eta, \x) B^i(\eta, \x)\right> = \int\frac{\d^3\k_1\,\d^3\k_2\,\d^3\k_3}{\left(2\pi\right)^{9/2}}e^{i (\k_1 + \k_2 + \k_3).\x}\left<\mathcal{R}_{\k_1}(\eta) B_{i\,\k_2}(\eta) B^i_{\k_3}(\eta)\right>,
\end{eqnarray}
where $B_i$ is related to the vector field $A_i$ as in \eqref{eq:define E B}. $\mathcal{R}_\k$ and $B_{i\,\k}$ are the Fourier modes associated with the curvature perturbation (see Eq. \eqref{eq:mode_curvature}) and i-$th$ component of the magnetic field, respectively.

In order to compute the higher-order correlation function during inflation, we adopt a very
useful and powerful tool of the in-in formalism \cite{Maldacena:2002vr}. In this formalism, the expectation value of an operator $\mathcal{O}$ at time $\eta$ is given by

\begin{equation}
	\left<\Omega\left|\mathcal{O}(\eta)\right|\Omega\right> = \left<0\left|\bar{T}\left(e^{i\int_{-\infty}^{\eta}H_{\rm int}(\eta^\prime)\d \eta^\prime}\right)\mathcal{O}(\eta)T\left(e^{-i\int_{-\infty}^{\eta}H_{\rm int}(\eta^{\prime\prime})\d \eta^{\prime\prime}}\right)\right|0\right>,
\end{equation}
where, $\left.|\Omega\right>$ and $\left.|0\right>$ are the vacuum of the interaction and free theories, respectively, $T$ and $\bar{T}$ are the time and anti time ordering operators, respectively, and $H_{\rm int}$ is the interaction Hamiltonian. Using leading order contribution of the interaction Hamiltonian, the cross-correlation function \emph{at the end of inflation, i.e.,  $\eta_{\rm end}$} can simply be written as \cite{Motta:2012rn, Jain:2012vm}
\begin{eqnarray}\label{eq:cross_corr}
	&&\left<\mathcal{R}_{\k_1}(\eta_{\rm end}) B_{i\,\k_2}(\eta_{\rm end}) B^i_{\k_3}(\eta_{\rm end})\right> = - i \int_{\eta_{\rm i}}^{\eta_{\rm end}} \d\eta \left<\left[\mathcal{R}_{\k_1}(\eta_{\rm end}) B_{i\,\k_2}(\eta_{\rm end}) B^i_{\k_3}(\eta_{\rm end}), H_{\rm int}(\eta)\right]\right>. \nonumber \\
&&	
\end{eqnarray}
Now, for simplicity, we define a new quantity $G_{\mathcal{R}BB}(\k_1, \k_2, \k_3)$ as
\begin{equation}\label{eq:cross_to_G}
	\left<\mathcal{R}_{\k_1}(\eta_{\rm end}) B_{i\,\k_2}(\eta_{\rm end}) B^i_{\k_3}(\eta_{\rm end})\right> \equiv \left(2\pi\right)^{-3/2} G_{\mathcal{R}BB}(\k_1, \k_2, \k_3)\delta^3(\k_1 + \k_2 +\k_3).
\end{equation}
Then, by using the expression \eqref{eq:cross_corr} along with the interaction Hamiltonian \eqref{eq:int_ham}, the definition of the mode functions \eqref{eq:mode_curvature} and \eqref{eq:Fourier expansion}, and the Wick's theorem,  $G_{\mathcal{R}BB}(\k_1, \k_2, \k_3)$ can be expressed as 
\begin{equation}
	G_{\mathcal{R}BB}(\k_1, \k_2, \k_3) = \sum_{i = 1}^{6}\mathcal{G}_i(\k_1, \k_2, \k_3),
\end{equation}
where six $\mathcal{G}$'s are defined as
\begin{eqnarray}
	\mathcal{G}_1(\k_1, \k_2, \k_3) &=& i\left(k_1^2 - k_2^2 -k_3^2\right)\mathcal{R}_{k_1}(\eta_{\rm end})A_{k_2}(\eta_{\rm end})A_{k_3}(\eta_{\rm end}) \times\nonumber \\
	&&\int^{\eta_{\rm end}}_{\eta_i} \d \eta\, f^2(\eta) \mathcal{R}^\ast_{k_1}(\eta)A^{\ast\prime}_{k_2} (\eta)A^{\ast\prime}_{k_3}(\eta)  + c.c. \\
	\mathcal{G}_2(\k_1, \k_2, \k_3) &=& -i\left(k_1^2 - k_2^2 -k_3^2\right)\mathcal{R}_{k_1}(\eta_{\rm end})A_{k_2}(\eta_{\rm end})A_{k_3}(\eta_{\rm end}) \times\nonumber \\
	&&\int^{\eta_{\rm end}}_{\eta_i} \d \eta\, \frac{f^2(\eta)}{a H} \mathcal{R}^{\ast\prime}_{k_1}(\eta)A^{\ast\prime}_{k_2} (\eta)A^{\ast\prime}_{k_3}(\eta)  + c.c.\\
	 \mathcal{G}_3(\k_1, \k_2, \k_3) &=& -i\left(k_2^2 k_3^2 + \frac{1}{4} \left(k_1^2 - k_2^2 -k_3^2\right)^2 \right) \mathcal{R}_{k_1}(\eta_{\rm end})A_{k_2}(\eta_{\rm end})A_{k_3}(\eta_{\rm end}) \times\nonumber \\
	&&\int^{\eta_{\rm end}}_{\eta_i} \d \eta\, f^2(\eta) \mathcal{R}^{\ast}_{k_1}(\eta) A^{\ast}_{k_2} (\eta) A^{\ast}_{k_3}(\eta)  + c.c. \\
	\mathcal{G}_4(\k_1, \k_2, \k_3) &=& i\left(k_2^2 k_3^2 - \frac{1}{4} \left(k_1^2 - k_2^2 -k_3^2\right)^2 \right) \mathcal{R}_{k_1}(\eta_{\rm end})A_{k_2}(\eta_{\rm end})A_{k_3}(\eta_{\rm end}) \times\nonumber \\
	&&\int^{\eta_{\rm end}}_{\eta_i} \d \eta \,\frac{f^2(\eta)}{a H} \mathcal{R}^{\ast\prime}_{k_1}(\eta) A^{\ast}_{k_2} (\eta) A^{\ast}_{k_3}(\eta)  + c.c.
\end{eqnarray}

\begin{eqnarray}
	\mathcal{G}_5(\k_1, \k_2, \k_3) &=& i\left(k_2^2 k_3^2 + \frac{1}{4} \left(k_1^2 - k_2^2 -k_3^2\right)\left(k_1^2 + 3 k_3^2 - k_2^2\right) \right) \mathcal{R}_{k_1}(\eta_{\rm end})A_{k_2}(\eta_{\rm end})A_{k_3}(\eta_{\rm end}) \times\nonumber \\
	&&\int^{\eta_{\rm end}}_{\eta_i} \d \eta\, \frac{f^2(\eta)}{a H} \mathcal{R}^{\ast}_{k_1}(\eta) A^{\ast\prime}_{k_2} (\eta) A^{\ast}_{k_3}(\eta)  + c.c.\\
	\mathcal{G}_6(\k_1, \k_2, \k_3) &=& i\left(k_2^2 k_3^2 + \frac{1}{4} \left(k_1^2 - k_2^2 -k_3^2\right)\left(k_1^2 + 3 k_3^2 - k_2^2\right) \right) \mathcal{R}_{k_1}(\eta_{\rm end})A_{k_2}(\eta_{\rm end})A_{k_3}(\eta_{\rm end}) \times\nonumber \\
	&&\int^{\eta_{\rm end}}_{\eta_i} \d \eta \,\frac{f^2(\eta)}{a H} \mathcal{R}^{\ast}_{k_1}(\eta) A^{\ast}_{k_2} (\eta) A^{\ast\prime}_{k_3}(\eta)  + c.c.
\end{eqnarray}
\noindent The abbreviated form of $c.c$ is complex conjugate. Now, given the coupling function $f(\eta)$, which fixes the solution of the vector field $A_\k(\eta)$, and the solution of the curvature perturbation $\mathcal{R}_\k$, one can easily compute the cross-correlations using the above expressions.
 
In our case, with the coupling function \eqref{eq:new choice coupling}, and the solution of the vector field $A_{\k}$, i.e., $$A_k(\eta) \simeq \frac{e^{-ik\eta}}{\sqrt{2\,k}} \left(d \left(\frac{\eta}{\eta_{\rm end}} \right)^5-1\right)\frac{\left(3 + 3ik\eta - k^2\eta^2\right)}{k^2\eta^2} $$ as well as the solution of the curvature perturbations $\mathcal{R}_\k$ \eqref{eq:mode_curvature}, we have evaluated all $\mathcal{G}$'s, and the explicit expressions are given in Appendix \ref{app:eval_G}. Please note that, in doing such evaluations, we assume that $d \ll 1$, which is evident from Sec. \ref{sec:solv_back}.

At this point, please notice the logarithmic dependence of $\mathcal{G}_3$ and $\mathcal{G}_4$.  Also, in all expressions of $\mathcal{G}$'s, $\eta_{\rm end}$ appears in the denominator. In $\mathcal{G}_3$ and $\mathcal{G}_4$, while it is $\eta_{\rm end}^4$, in other four $\mathcal{G}$'s, it appears as $\eta_{\rm end}^6$. Later, while evaluating the non-Gaussianity parameter in the next section, we will show that, while $\eta_{\rm end}^4$ dependence cancels out, $\eta_{\rm end}^6$ as well as the logarithmic dependencies, in general, do not. And therefore, for $\eta_{\rm end}\rightarrow0$ limit, except for the case of squeezed limit, it contributes enormously in evaluating the non-Gaussianity parameter. 

\section{Non-Gaussianity parameter}\label{sec:ng}

In the last section, we have defined the cross-correlation function between the magnetic fields and the curvature perturbation. However, it is not dimensionless, and therefore, to characterize the correlation function more prominently, we introduce the non-Gaussianity parameter $b_{\rm NL}$. The parameter is dimensionless and captures the shape and amplitude of the three-point cross-correlation function discussed in the previous section.

The non-Gaussianity parameter $b_{\rm NL}$ in the local form is defined as 

\begin{eqnarray}
\hat{B}_{i\,\bf k}(\eta) 
= \hat{B}_{i\,\bf{k}}^{({\rm G})}(\eta) 
+ b_{\rm NL}\, 
\int \frac{\d^3{\bf k^\prime}}{(2\,\pi)^{3/2}}\,
\hat{\mathcal{R}}^{(\rm G)}_{\bf{k}-\bf{k^\prime}}(\eta)\,
\hat{B}_{i\,{\bf k^\prime}}^{({\rm G})}(\eta),
\end{eqnarray}

\noindent where $\hat{B}_{i\,\bf k}(\eta)$ is the Fourier model of the magnetic field and $(G)$
denotes the Gaussian part of the Fourier mode. Using the above expression, one can now evaluate the cross-correlation function \eqref{eq:cross_to_G} that interests us and then compare with the same that leads to

\begin{equation}
	b_{\rm NL}(\k_1, \k_2, \k_3) = -\frac{f^2(\eta_{\rm end})}{8 \pi^4 a^4(\eta_{\rm end})}\frac{k_1^3 k_2^3 k_3^3\, G_{\mathcal{R}BB}(\k_1, \k_2, \k_3)}{\mathcal{P}_{\mathcal{R}}(\k_1)\left(k_3^2 \mathcal{P}_{\rm B}(\k_2) + k_2^2 \mathcal{P}_{\rm B}(\k_3)\right)},
\end{equation}

\noindent where, $\mathcal{P}_{\rm B}$ is defined in $\eqref{eq:MagSpec}$ and $\mathcal{P}_{\mathcal{R}(\k)}$ is the power spectrum of the curvature perturbation, defined as  $$\mathcal{P}_{\mathcal{R}}(\k) \equiv \frac{k^3}{2\pi^2} \left|\mathcal{R}_{\k}\right|^2.$$ At the end of inflation, both the spectra, in our case,  take the form

\begin{equation}
	\mathcal{P}_{\mathcal{R}} (\k) \approx \frac{H^2}{8 \pi^2\,\epsilon} \qquad \mathcal{P}_{\rm B}(\k) \approx \frac{9 H^2}{4 \pi^2}.
\end{equation}
\noindent With the help of the above spectra, along with the six $\mathcal{G}$'s given in Appendix \ref{app:eval_G}, one can easily evaluate $b_{\rm NL}$. Since, the expression is enormous, as obvious from the Appendix \ref{app:eval_G}, one can simply obtain the same in the three interesting limits: squeezed, equilateral and flattened and study it extensively.

\subsection{Squeezed limit}
In the squeezed limit, one of the wavenumber is significantly smaller than the other two. Since the wavevctors form a triangle configuration as $\k_1 + \k_2 + \k_3 = 0$, $\k_1 \rightarrow 0$ implies that $\k_2 = - \k_3$. In this limit, the non-Gaussianity parameter becomes:

\begin{eqnarray}
	\lim\limits_{\k_1 \rightarrow 0}b^{\rm sq}_{\rm NL}(\k_1, \k, -\k) &=& \lim\limits_{-k\eta_{\rm end}\rightarrow 0}\,\frac{3(6 + k^2 \eta_{\rm end}^2)}{9 + 3 k^2 \eta_{\rm end}^2 + k^4 \eta_{\rm end}^4} = 2.
\end{eqnarray}

\noindent Two things to note: $b_{\rm NL}$ is independent of $d$ and $k$. The result is not surprising as it is expected because of the consistency condition which is satisfied during inflation \cite{Jain:2012ga, Jain:2012vm}. It tells us that, in the squeezed limit, the non-Gaussianity parameter can be written as the spectral index of the magnetic field energy density  as $$b^{\rm sq}_{\rm NL}(\k_1, \k, -\k) = \frac{4 - n_{\rm B}}{2},$$ where, $$n_{\rm B} \equiv \frac{\d \ln \mathcal{P}_{\rm B}}{\d \ln k}.$$ Since, the choice of the coupling function \eqref{eq:new choice coupling} leads to scale-invariant magnetic field spectrum, the corresponding spectral index $n_{\rm B}$ is zero, which makes $b_{\rm NL}$ to be equal to two.

The relation has previously been obtained with the help of power-law kind of coupling function, i.e., $f \propto (-\eta)^\alpha$ \cite{Ferreira:2013sqa, Ferreira:2014hma}. However, although the magnetic spectrum is scale-invariant, our proposed coupling function is not power-law, and even after that, with this new choice \eqref{eq:new choice coupling}, we obtain the result: \emph{consistency relation is satisfied}. The result is interesting and unique as in the next section, it will be apparent that in the other limits, the non-Gaussianity parameter depends both on $k$ and $d$. This is another interesting result of this work.

\subsection{Equilateral limit}

Unlike the squeezed limit, in the equilateral limit, the non-Gaussianity parameter behaves differently and possesses unique signatures. In this limit, as it forms equilateral triangle, it implies $k_1 = k_2 = k_3$. In this configuration, the non-Gaussianity parameter acts as

\begin{eqnarray}\label{eq:bnleq}
	b^{\rm eq}_{\rm NL}(k, k, k) = 4.11 - \frac{2.37\,d}{k^2\eta_{\rm end}^2}- 2.62 \log|k\eta_{\rm end}|.
\end{eqnarray}

\noindent Such form of non-Gaussianity parameter is unique: it contains two kinds of $\eta_{\rm end}$ dependence. One kind appears with the parameter $d$ in the denominator, while the other enters into the logarithmic. The logarithmic dependence is natural and explored before for $f \propto (-\eta)^{-2}$, while the former one is unexplored to date.

\begin{figure}
	\centering
	\includegraphics[scale=.37]{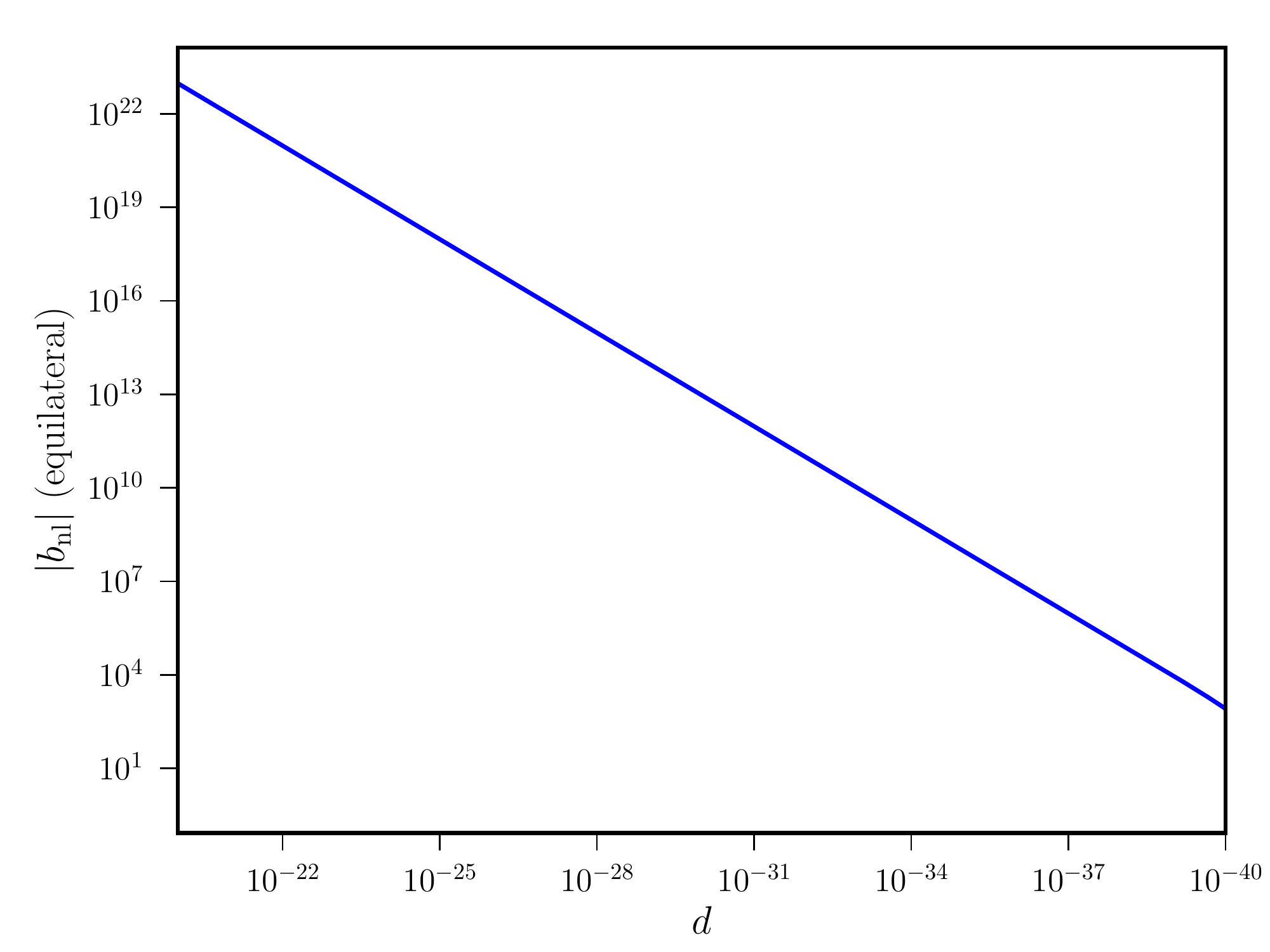}
	\includegraphics[scale=.37]{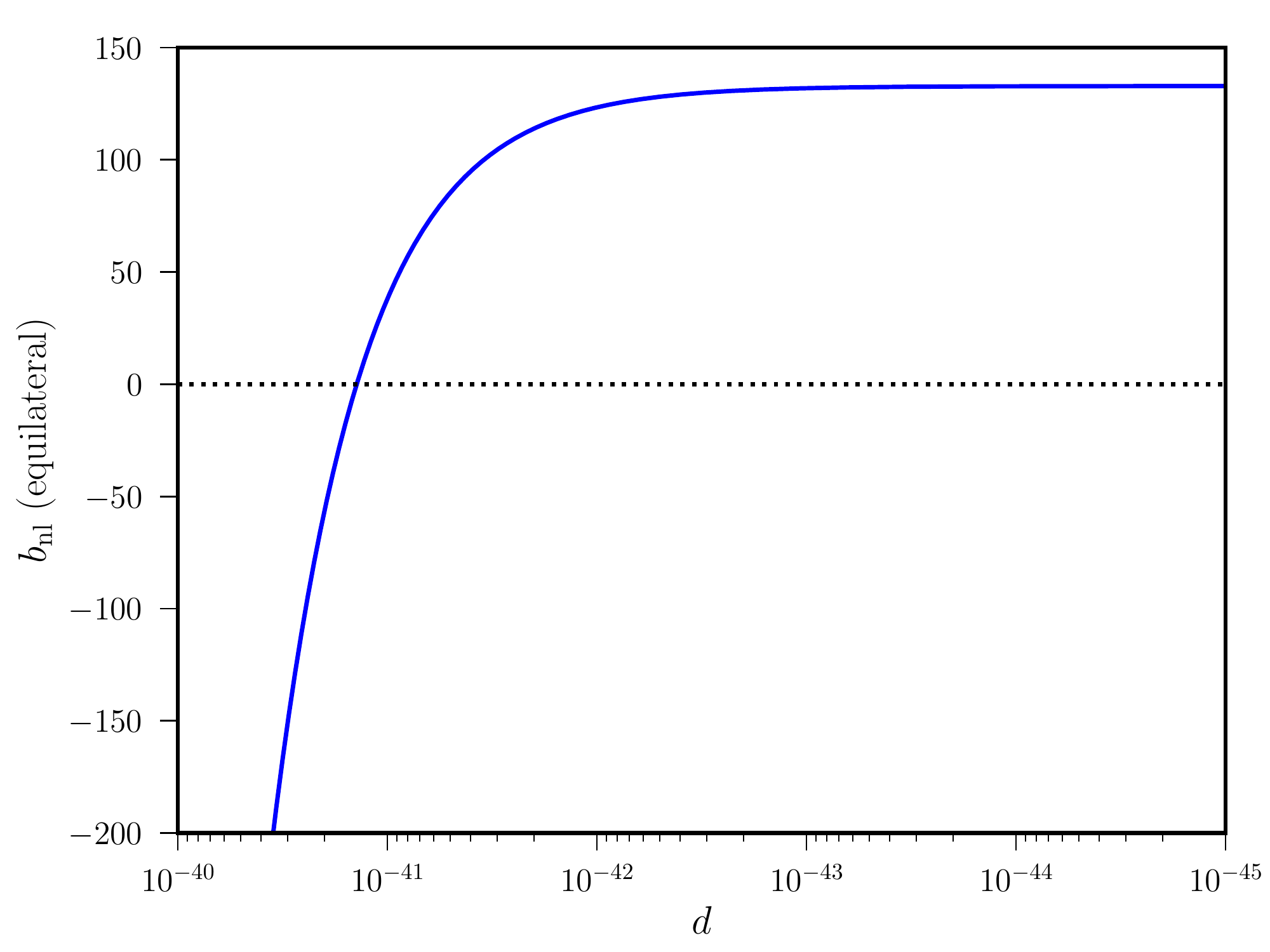}
	\caption{\label{fig:bnleqd}$b^{\rm eq}_{\rm NL}$ is plotted against $d$ for the pivot scale $k = 0.05$ Mpc. For $d > 10^{-41},$ it behaves as $\propto d$, whereas, for $d < 10^{-41}$, it quickly saturates at $132.$}
\end{figure}

The above expression also provides us an interesting scenario: for a given wavenumber, $b_{\rm NL}$ vanishes at  $d_\ast \simeq - k^2 \eta_{\rm end}^2 \log|k\eta_{\rm end}|$. When $d > d_\ast$, the non-Gaussian parameter becomes negative and entirely behaves as $$ b^{\rm eq}_{\rm NL} \simeq - \frac{2.37\,d}{k^2\eta_{\rm end}^2}.$$ For $ d > d_\ast,$ it becomes positive and independent of $d$, and quickly reaches to the saturated value $$ b^{\rm eq}_{\rm NL} \simeq - 2.62\, \log|k\eta_{\rm end}|.$$ This is plotted in Fig. \ref{fig:bnleqd}.

\begin{figure}
	\centering
	\includegraphics[scale=.37]{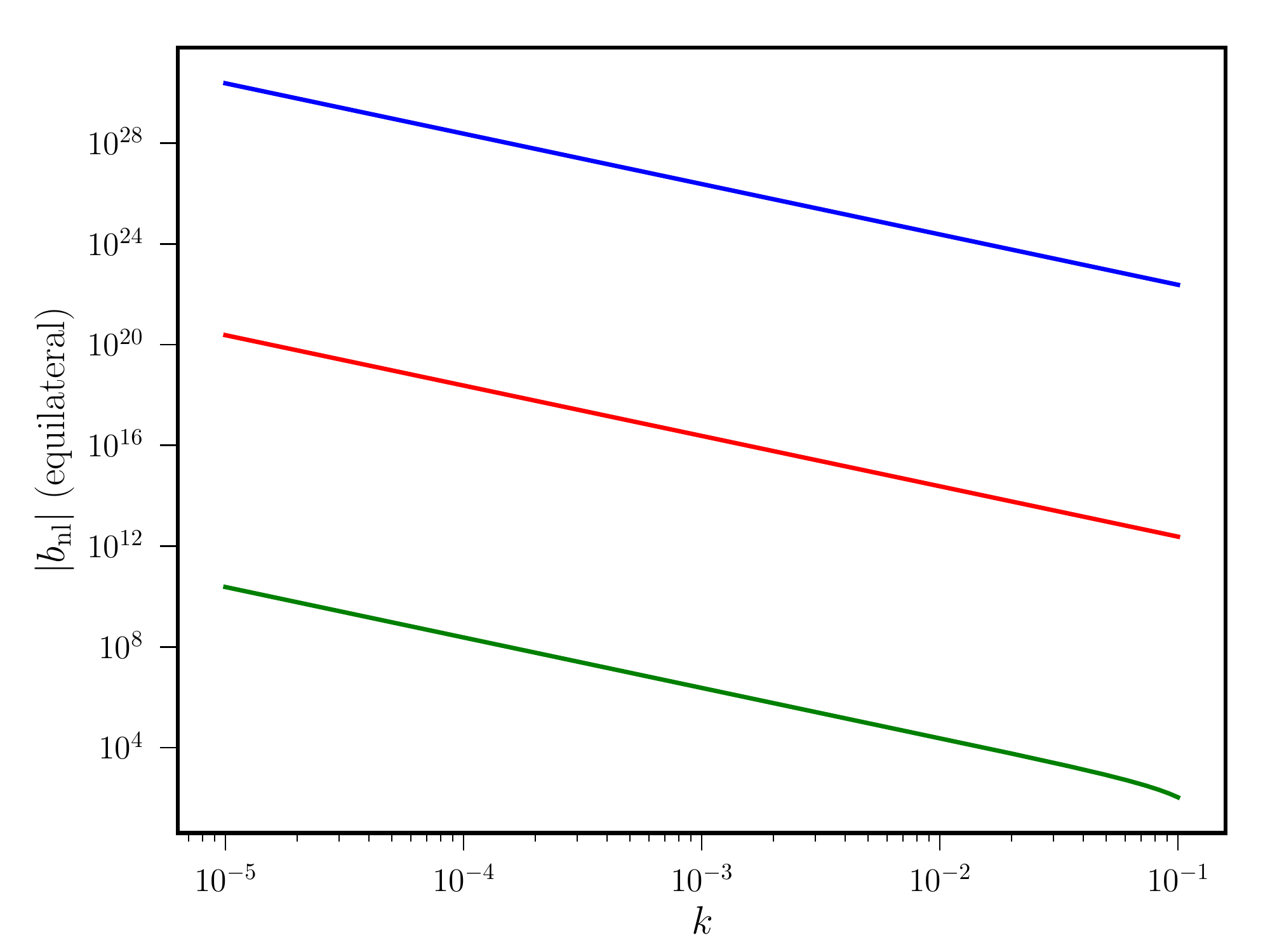}
	\includegraphics[scale=.37]{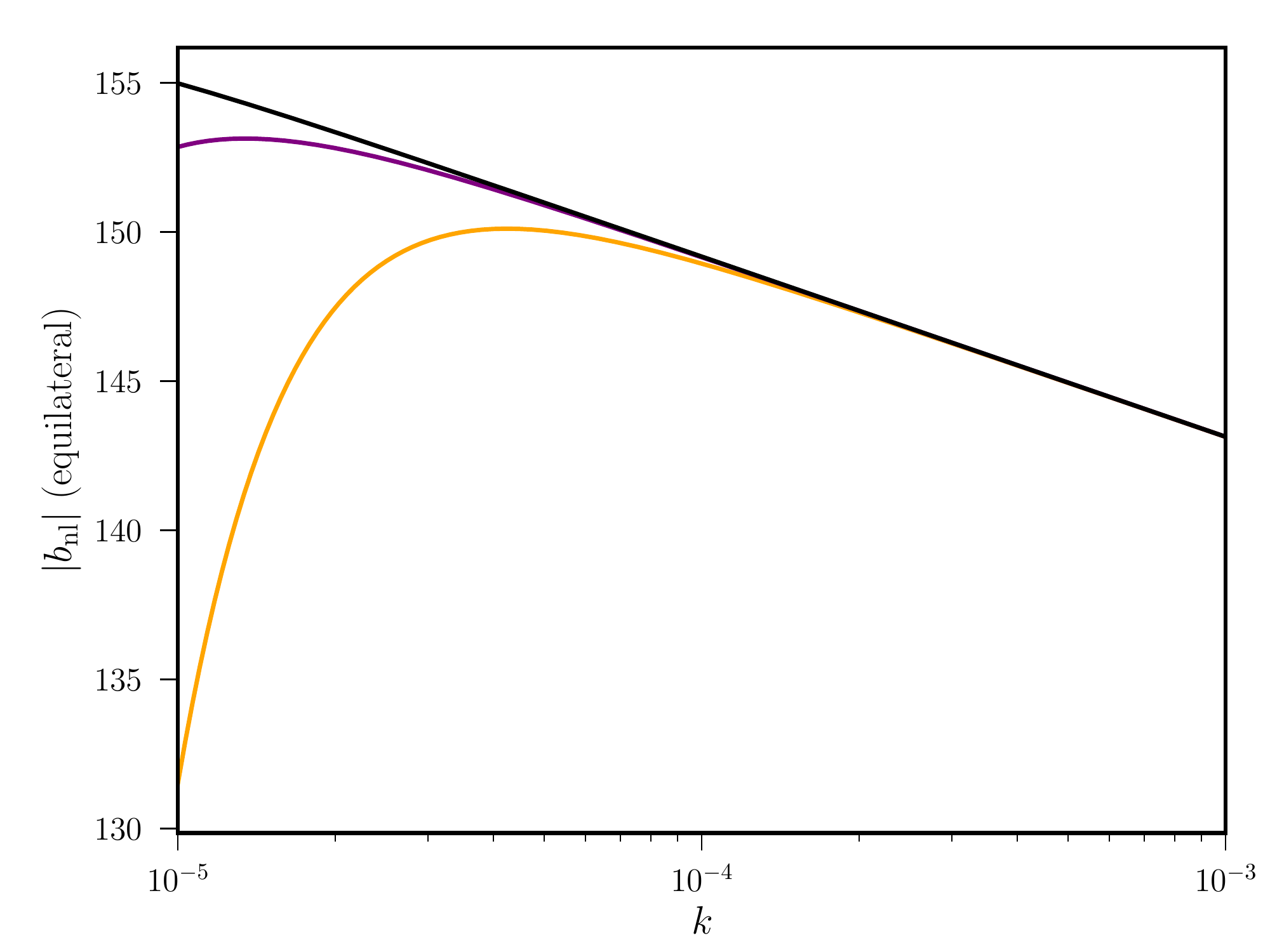}
	\caption{\label{fig:bnleqk} We plot $b^{\rm eq}_{NL}$ against $k$ for different values of $d$ (blue for $10^{-20},$ red for $d = 10^{-30},$ green for $d = 10^{-40}$, orange for $d = 10^{-49}$, purple for $d = 10^{-50}$ and black for $10^{-51}$).}
\end{figure}

\begin{figure}
	\centering
	\includegraphics[scale=.37]{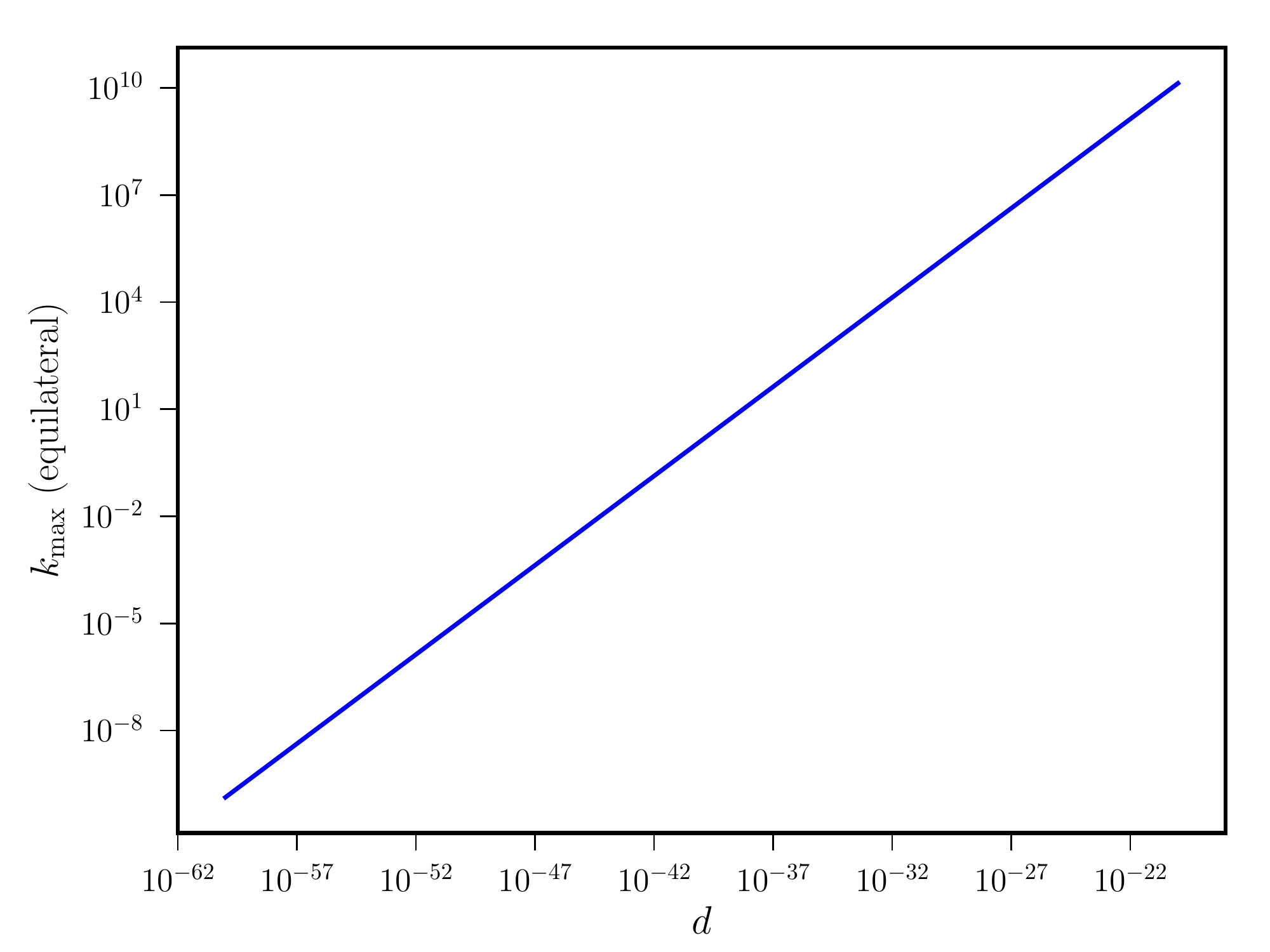}
	\includegraphics[scale=.37]{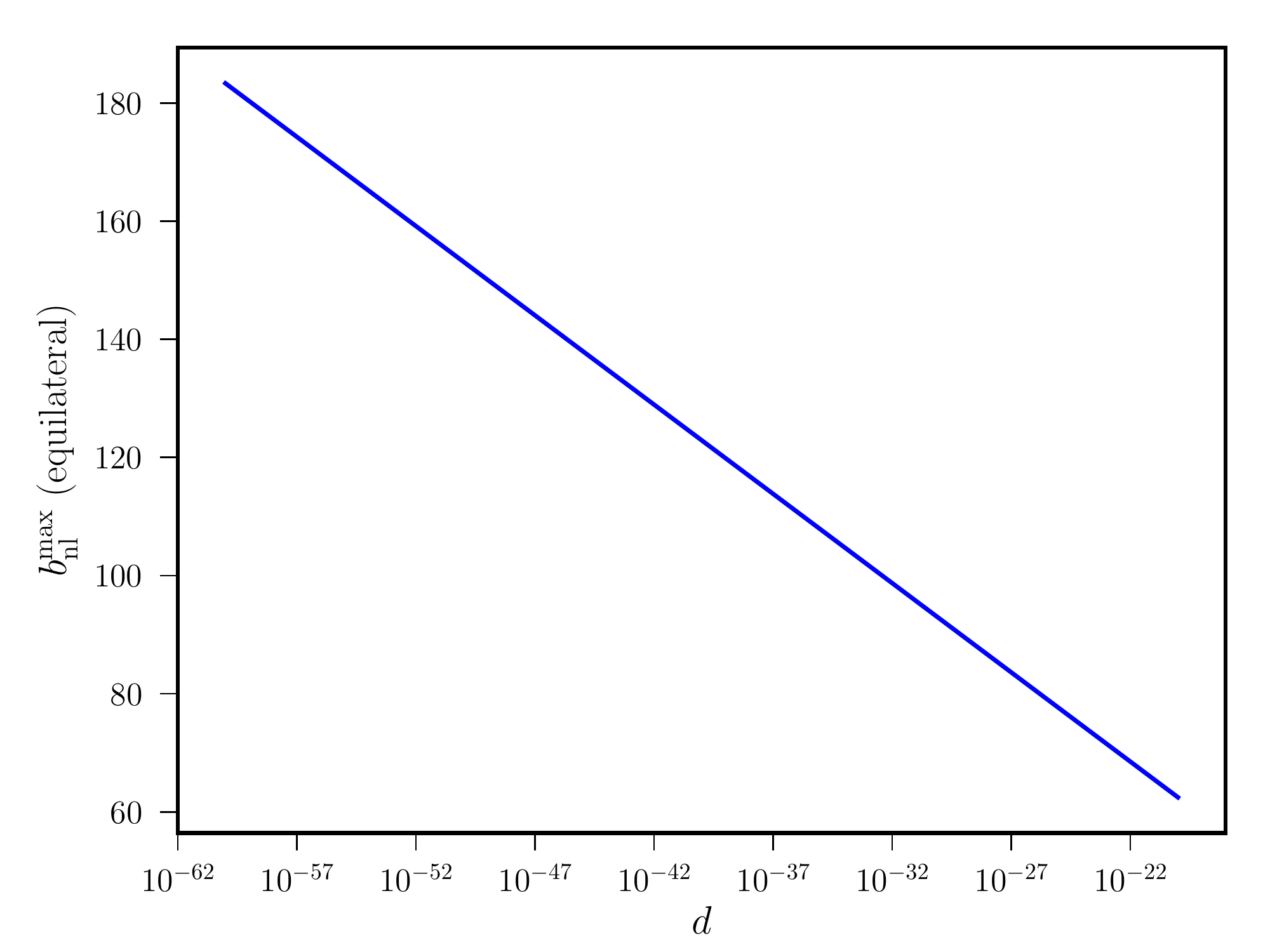}
	\caption{\label{fig:eqmax}In the left, we plot $k_{\rm max}\propto \sqrt{d}$ against $d$ in the equilateral limit. In the right, we plot $b^{\rm max}_{\rm NL}$ against $d$. As one can see, the maximum value of the non-Gaussianity parameter decreases with $d$.}
\end{figure}

In Fig. \ref{fig:bnleqk}, for different values of $d$, we plot the $b_{\rm NL}$ in the equilateral limit. As obvious from \eqref{eq:bnleq}, for a given value of $k$ such that $d > - k^2 \eta_{\rm end}^2 \log|k\eta_{\rm end}|,$ then the non-Gaussianity parameter behaves as $\propto k^{-2},$ whereas, for $d > - k^2 \eta_{\rm end}^2 \log|k\eta_{\rm end}|,$ it becomes positive and possesses a maximum at $$k_{\rm max} = -\frac{1.34 \sqrt{d}}{\eta_{\rm end}}, \quad b^{\rm max}_{\rm NL} = 2.79 + 0.608\,d - 2.62\,\log[1.34\,\sqrt{d}].$$ After that, it becomes independent of $d$ and behaves logarithmic. In Fig. \ref{fig:eqmax}, $k_{\rm max}$ and $b^{\rm max}_{\rm NL}$ is plotted against $d$. It becomes apparent that, in equilateral limit, $b^{\rm max}_{\rm NL} \sim \mathcal{O}(2).$

From these figures, it is now obvious that the \emph{in the equilateral limit, the non-Gaussianity parameter corresponding to the pivot scale can be as large as $10^{23}$ to zero, depending on the value of the model parameter $d$, which is huge and never explored.}

\subsection{Flattened limit}

\begin{figure}
	\centering
	\includegraphics[scale=.37]{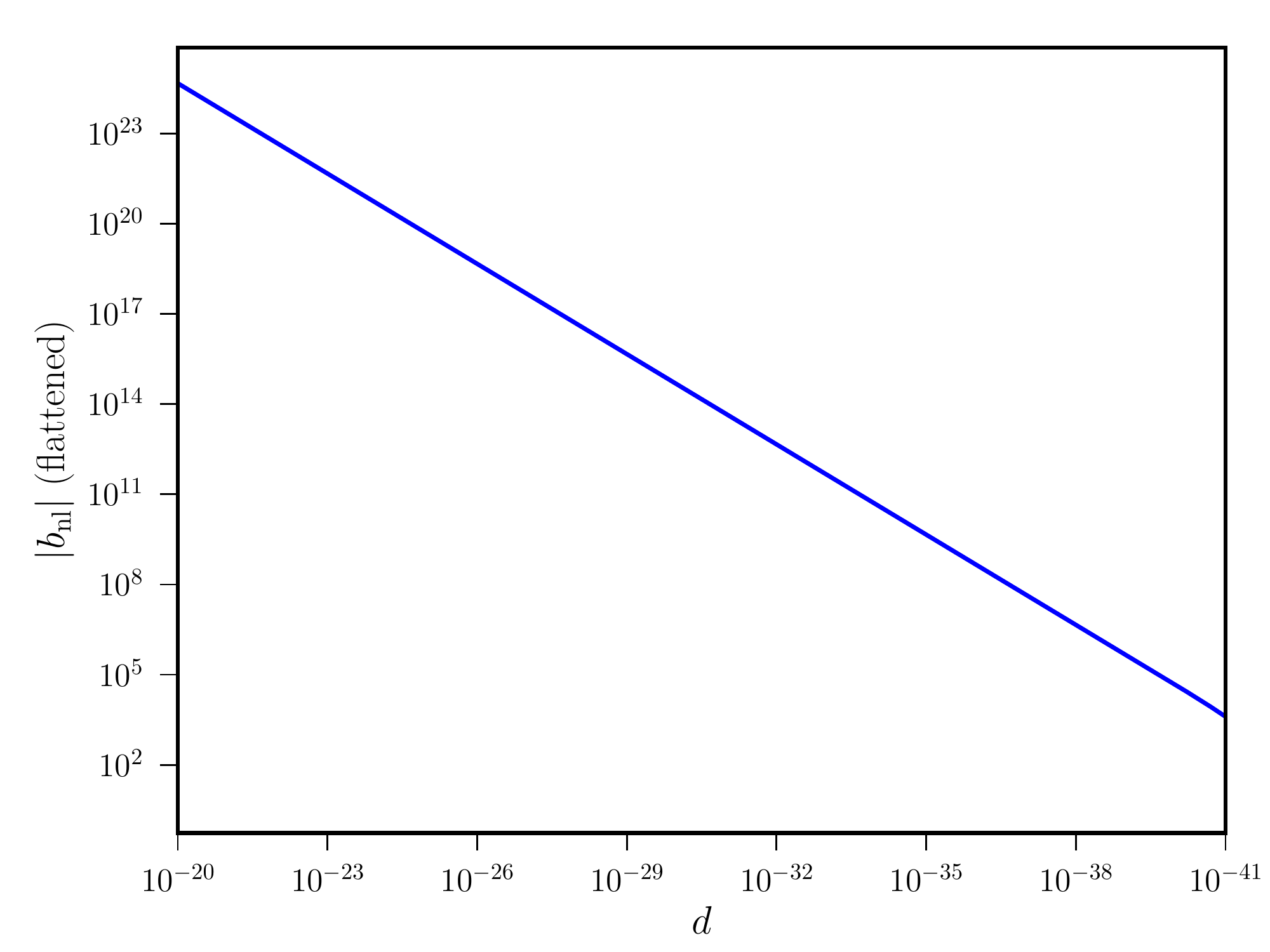}
	\includegraphics[scale=.37]{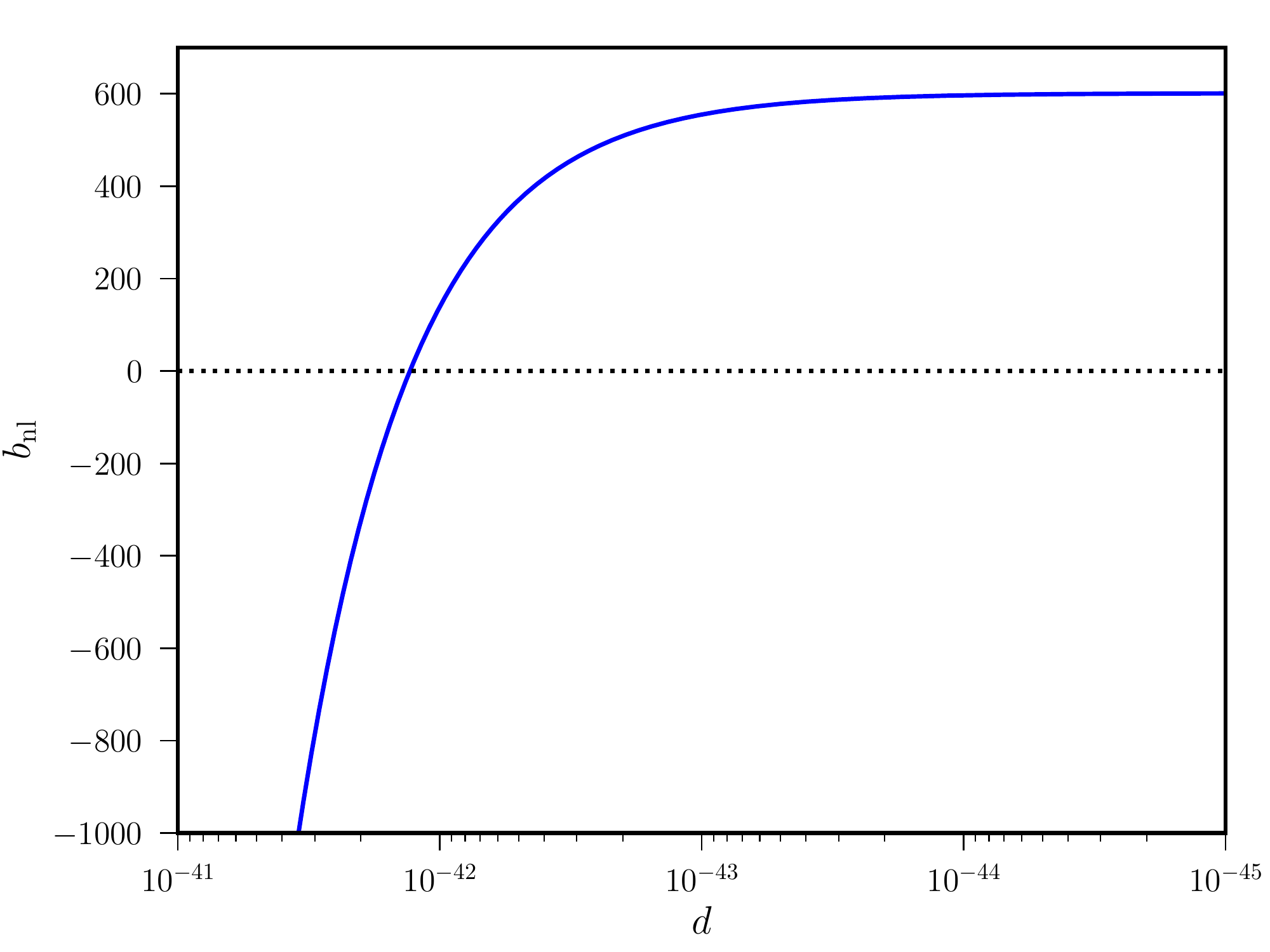}
	\caption{\label{fig:bnlfld}$b^{\rm fl}_{\rm NL}$ is plotted against $d$ for the pivot scale $k = 0.05$ Mpc. For $d > 10^{-41},$ it behaves as $\propto d$, whereas, for $d < 10^{-41}$, it quickly saturates at $\sim 600.$}
\end{figure}

\begin{figure}
	\centering
	\includegraphics[scale=.37]{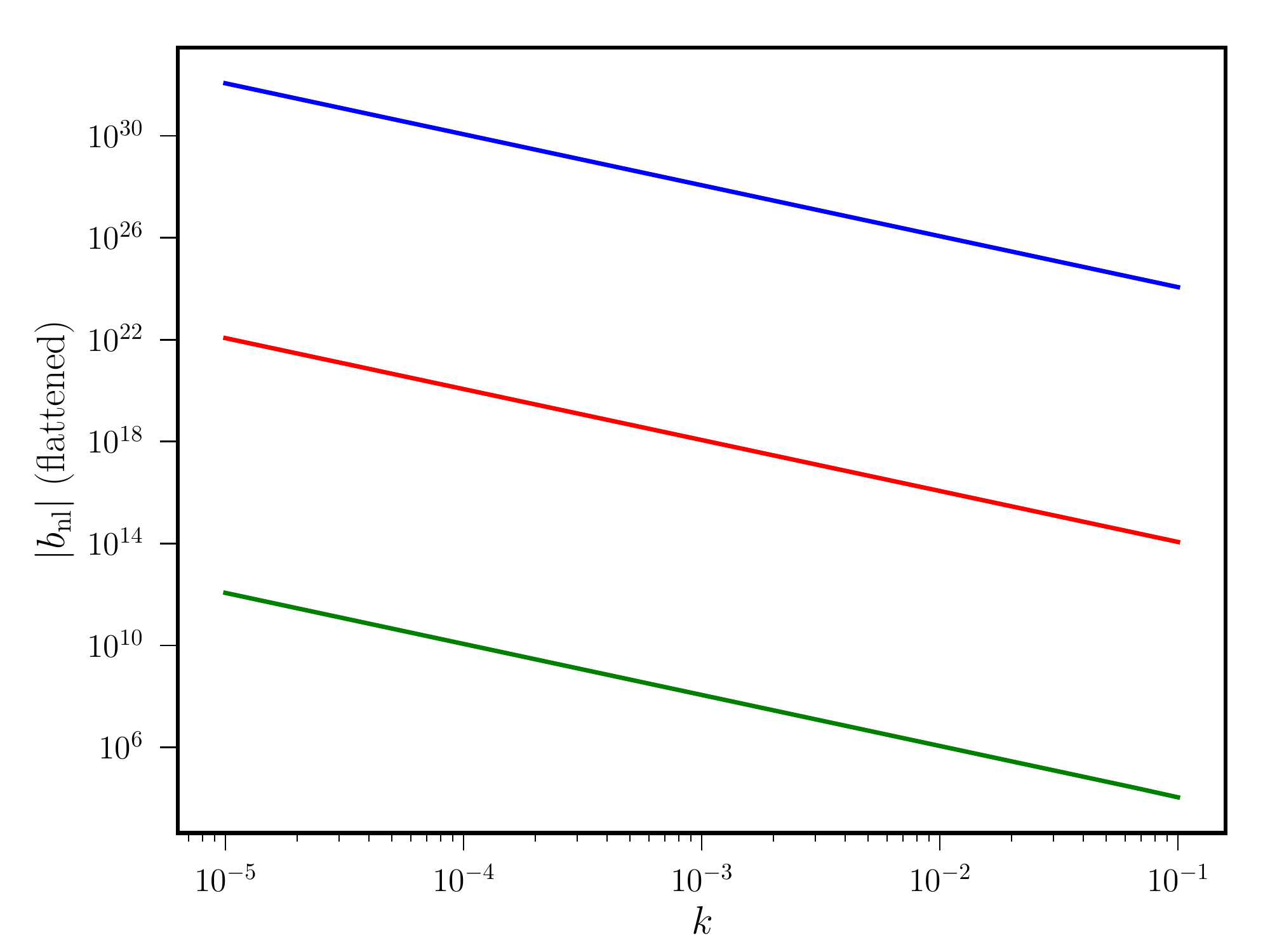}
	\includegraphics[scale=.37]{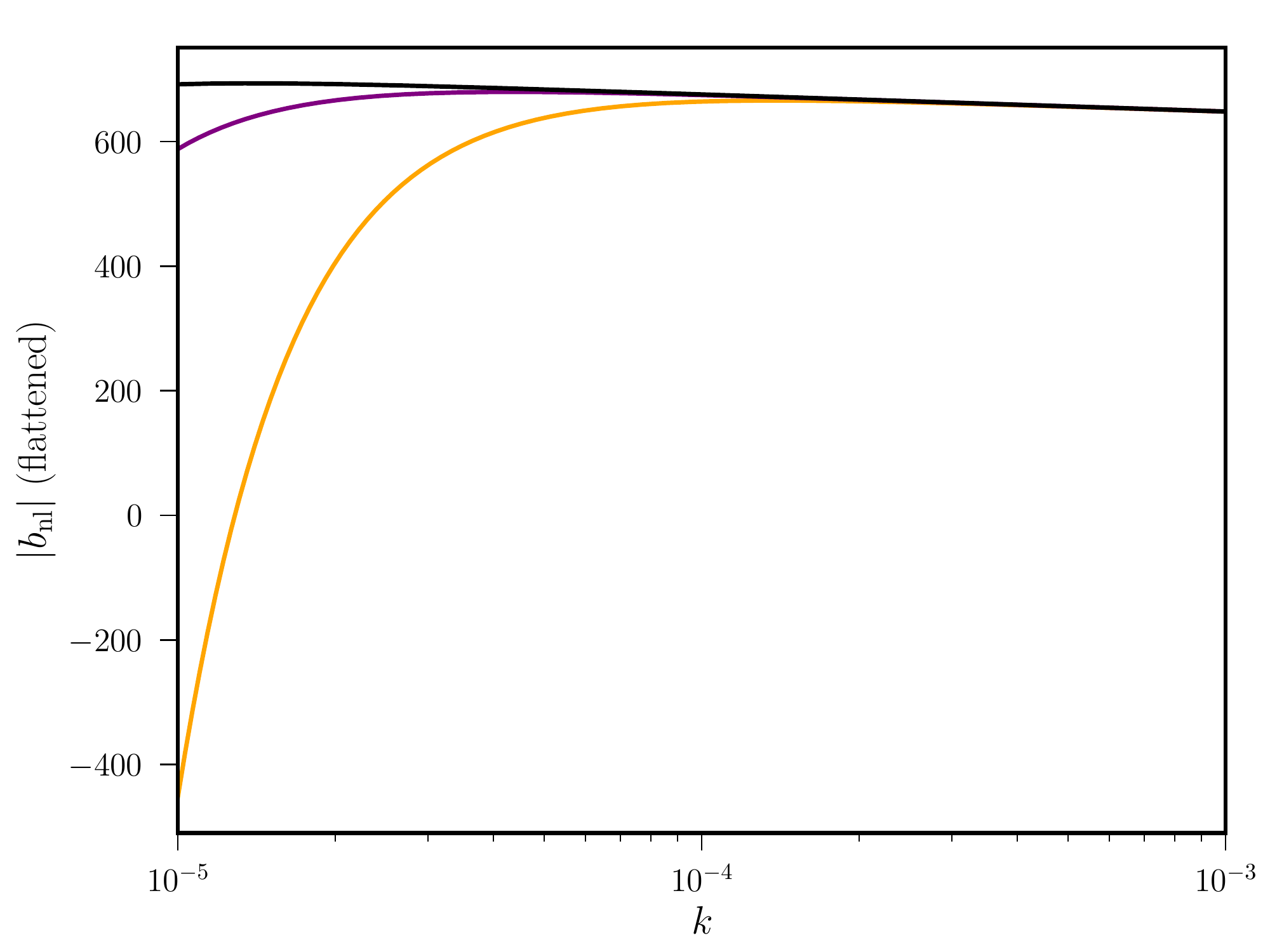}
	\caption{\label{fig:bnlflk} We plot $b^{\rm fl}_{NL}$ against $k$ for different values of $d$ (blue for $10^{-20},$ red for $d = 10^{-30},$ green for $d = 10^{-40}$, orange for $d = 10^{-49}$, purple for $d = 10^{-50}$ and black for $10^{-51}$).}
\end{figure}

The flattened limit implies that $k_1 = 2 k_2 = 2 k_3$, and in this limit, similar to equilateral limit, $b_{\rm NL}$ behaves as

\begin{equation}
	b^{\rm fl}_{\rm NL} = 12.7-\frac{115.6\,d}{k^2\eta_{\rm end}^2} - 12\log|k\eta_{\rm end}|
\end{equation}
\noindent Again, similar to equilateral limit, for $d = d_\ast \simeq -0.1\,k^2\eta_{\rm end}^2\,\log|k\eta_{\rm end}|$, the non-Gaussianity parameter vanishes. For $d$ greater than that value, it behaves as $$ -\frac{115.6\,d}{k^2\eta_{\rm end}^2},$$ while for $d < d_\ast$, it behaves as $$- 12\log|k\eta_{\rm end}|.$$ Such dependencies of $b^{\rm fl}_{\rm NL}$ have been plotted in Fig. \ref{fig:bnlfld} and Fig. \ref{fig:bnlflk}. Also, $k_{\rm max}$ and $b^{\rm max}_{\rm NL}$ for this case yield as

$$k_{\rm max} \simeq -4.39\frac{\sqrt{d}}{\eta_{\rm end}}, \quad b^{\rm max}_{\rm L} \simeq 6.75 - 40.67\,d - 12\log[4.38 \sqrt{d}].$$

\begin{figure}[!h]
	\centering
	\includegraphics[scale=.37]{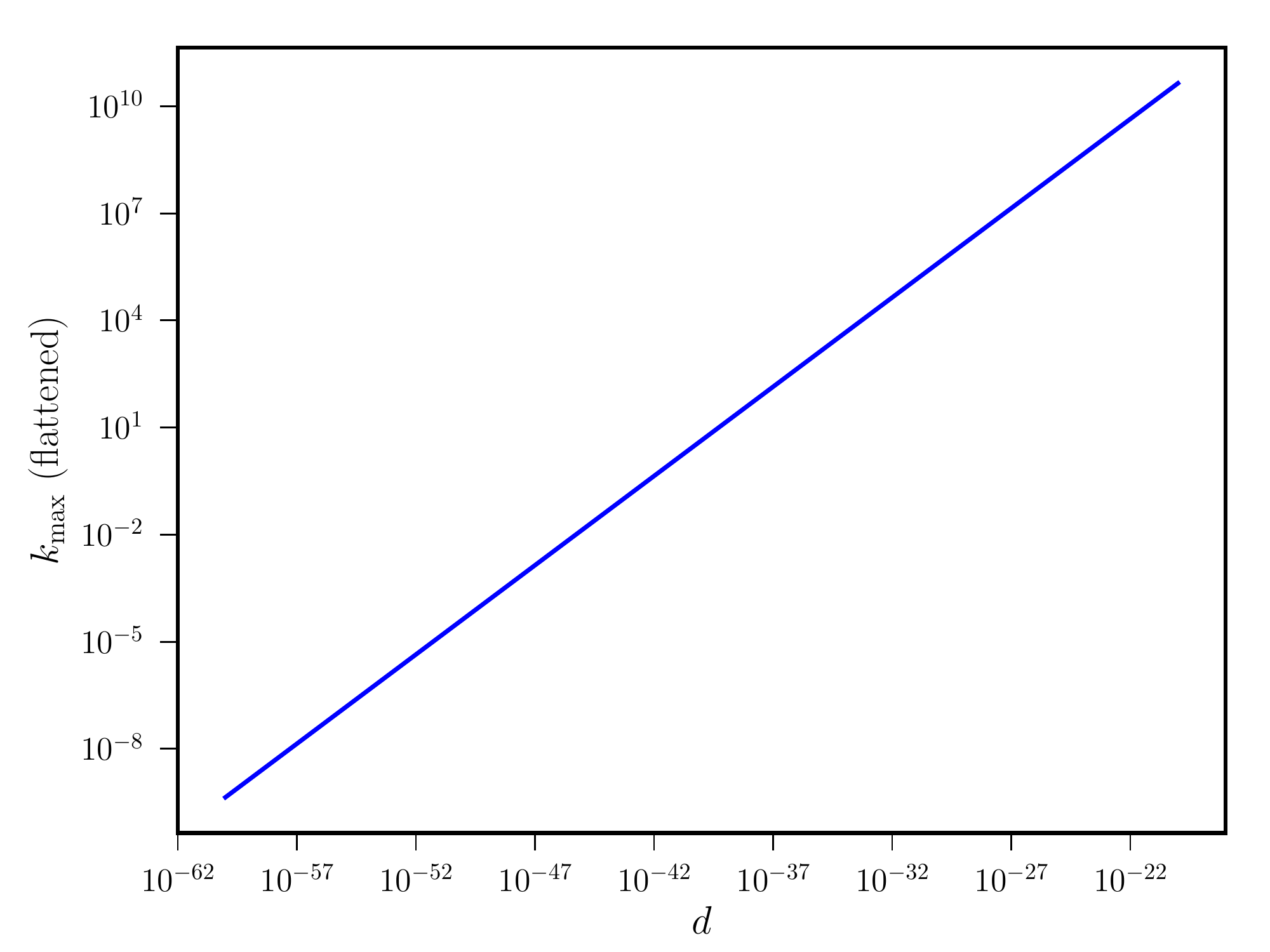}
	\includegraphics[scale=.37]{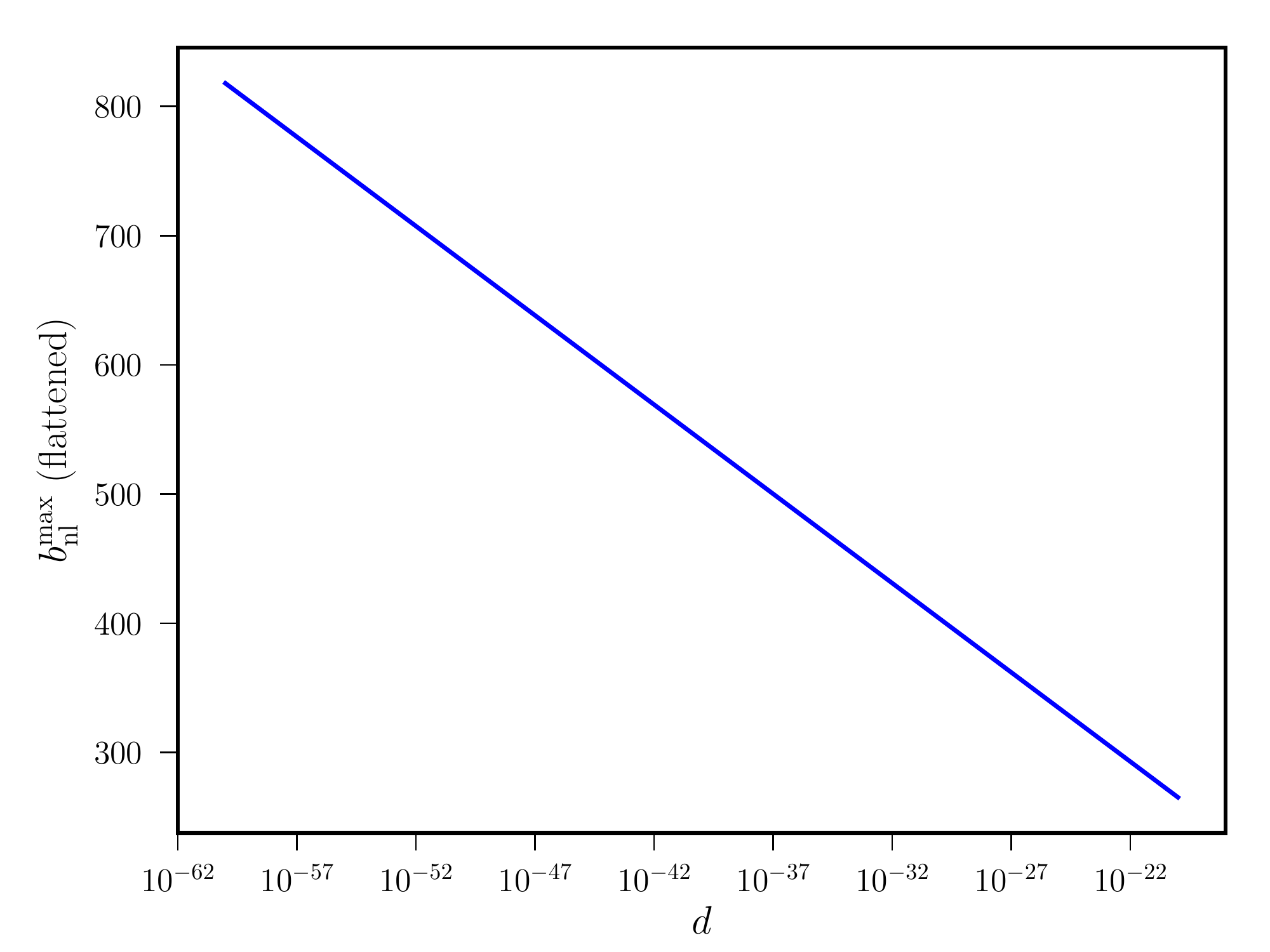}
	\caption{\label{fig:flmax}In the left, we plot $k_{\rm max}\propto \sqrt{d}$ against $d$ in the equilateral limit. In the right, we plot $b^{\rm max}_{\rm NL}$ against $d$. As one can see, the maximum value of the non-Gaussianity parameter decreases with $d$.}
\end{figure}

\noindent These have been plotted in Fig. \ref{fig:flmax}. From these figure, it becomes apparent that the non-Gaussianity parameter in flattened limit always generate higher value than that in the equilateral limit.  For $d > d_\ast$,  the ratio between them, i.e., $b^{\rm fl}_{\rm NL}/b^{\rm eq}_{\rm NL} \sim 50$, whereas, for $d < d_\ast,$ it is $\sim 4.6$.

Similar to the equilateral limit, it is now obvious that the \emph{in the flattened limit, the non-Gaussianity parameter corresponding to the pivot scale can be as large as $10^{24}$ to zero, depending on the value of the model parameter $d$. This is another interesting result of our work.}

\section{Conclusions}\label{sec:conclu}

Inflationary magnetogenesis has long been studied in the literature that can explain the current magnetic field abundances in the cosmic scales. In doing so, many mechanisms have been proposed and among them, $-1/4 f^2(\phi) F_{\mu \nu} F^{\mu \nu}$ model grabs the most attention. At present time, it can safely generate $10^{-13}$ G at Mpc length scale. For two kinds of coupling function solutions, this can be achieved. However, both of them suffer severe issues: while one leads to high electric field energy density that can spoil the background and the model cannot self-sustain, another leads to what is referred to as the strong coupling problem. In this case, the coupling function, at the beginning of inflation possesses a very tiny value $\ll 1$, which implies that the effective coupling constant is extremely high, i.e., $\gg 1$ which, in turn, makes the quantization process of the vector field theory untrustworthy. Due to these issues, one can show that such model can generate not more than $10^{-30}$ G magnetic field self-consistently in the Mpc length scales \cite{Demozzi:2009fu}.

We noticed that, although each of the scale-invariant solutions suffers from backreaction or the strong coupling, each of them also solves the other problem, i.e., the solution with strong coupling does not possess backreaction and vice-versa. Motivated by this, we proposed a new coupling function which is a linear combination of both the coupling functions. In doing so, we introduce a model parameter that characterizes the linear combination and showed that, in order to evade the backreaction problem, one finds the value to be extremely small i.e., $d \lesssim 10^{-20}.$ We also found that, for such model with a non-zero value of $d$, it contains two distinct regions:

\begin{itemize}
	\item[1.] for $|\eta| \gtrsim d^{-1/3} |\eta_{\rm end}|,$ there is no strong coupling,
	\item[2.] for $|\eta_{\rm end}| \lesssim |\eta| \lesssim d^{-1/3}\,|\eta_{\rm end}|,$ the model enters into the strong coupling region.
\end{itemize}

\noindent Assuming the pivot scale $k = 0.05$ Mpc${}^{-1}$ exiting the Hubble horizon 50 e-folds before the end of inflation, we find that, for $10^{-20} \lesssim d \lesssim 10^{-60}$, the modes corresponding to the Mpc length scales become super-Horizon and \emph{classical}, similar to inflationary perturbations before the system enters into the strong coupling region. Therefore, \emph{one can argue that, while very short wavelength modes suffer the problem of strong coupling, the high wavelength modes that we are interested in do not suffer the same, thus solving the problem of strong coupling for primordial magnetogenesis.} In this way, we showed that, today, up to $10^{-13}$ G strength of the magnetic field can be generated self-consistently in the Mpc scales, which earlier was only $\sim 10^{-30}$ G \cite{Vernizzi:2004nc, Ferreira:2013sqa, Ferreira:2014hma}.

We, further extended our work to find the non-Gaussian signature of this model. We found that the model satisfies the consistency relation, i.e., in the squeezed limit, the non-Gaussianity parameter can be written in terms of the magnetic field spectra index as $b_{\rm NL} = (4 - n)/2.$  We also found the unique signatures of $b_{\rm NL}$ in the equilateral and flattened limit as:

$$b_{\rm NL}(k) \sim -C_1\frac{d}{k^2\eta_{\rm end}^2} - C_2 \log|k \eta_{\rm end}|.$$ For a certain range of $k$, while it is negative and  $ \propto k^{-2}$, for the rest, it becomes positive and $ \propto \log|k\eta_{\rm end}|$ with the flattened limit producing greater amplitude than the same in equilateral limit. This implies that \emph{the non-Gaussianity parameter can be as high as $10^{20}$ and as small as $\mathcal{O}(1)$ depending on the model parameter $d$.} These unique signatures can easily be verified from the future experiments which can detect these kinds of non-linearity.

The scope of such model is huge. One immediate extension is to study the effect of similar coupling functions in the context of helical magnetic field generation, which has been studied with the help of the auxiliary scalar field in Ref. \cite{Chowdhury:2018mhj}. Inflationary Universe, too, suffers from many issues, for which, there is a growing interest in finding alternatives to inflation and among them, the classical bouncing scenario is the most popular one. It would also thus be interesting to investigate the effect of such coupling function in the bouncing scenario. Recently, in Ref. \cite{Chowdhury:2018blx}, it has been extensively studied with the help of simple auxiliary scalar field coupling of $f \propto a(\eta)^\alpha.$ The auxiliary field has been used due to the fact that the curvature perturbations in the bouncing model do not satisfy the observational constraint. However, in Refs. \cite{Nandi:2020sif, Nandi:2020szp}, with the help of non-minimal coupling, the first viable bouncing model has been constructed (also see Refs. \cite{Nandi:2018ooh,Nandi:2019xag, Nandi:2019xlj} in these regard). These works are now in progress.

\section*{Acknowledgements}

The author wishes to thank L. Sriramkumar and Kinjalk Lochan for useful discussion. The author also thanks Sagarika Tripathy for the help in evaluating the non-Gaussian parameter.

\appendix

\section{Evaluating the $\mathcal{G}$'s}\label{app:eval_G}
\begin{eqnarray}
	\mathcal{G}_1(\k_1, \k_2, \k_3) &=& -\frac{H^2}{8\,\epsilon \,k_1^3\, k_2^5\, k_3^5\, k_T^6\, \eta_{\rm end}^6} (k_1^2 - k_2^2 - 
	k_3^2) (k_2^2 k_3^2 k_T^4 \eta_{\rm end}^2  (-9 (k_1^3 + 2 k_1^2 (k_2 + k_3) + \nonumber \\
	&&  
	2 k_1 (k_2^2 + k_2 k_3 + k_3^2) + (k_2 + k_3) (k_2^2 + k_2 k_3 + 
	k_3^2))  -  
	3 (2 k_1 k_2^2 k_3^2 + \nonumber \\
	&&k_2^2 k_3^2 (k_2 + k_3) + 
	2 k_1^3 (k_2^2 + k_3^2) + 
	2 k_1^2 (k_2 + k_3) (k_2^2 + k_2 k_3 + k_3^2)) \eta_{\rm end}^2 +\nonumber \\
	&& 
	k_2^2 k_3^2 (-4 k_1^3 + 2 k_1 k_2 k_3 - k_1^2 (k_2 + k_3) + 
	k_2 k_3 (k_2 + k_3)) \eta_{\rm end}^4 + \nonumber \\
	&&
	k_1^2 k_2^3 k_3^3 (k_1 + k_2 + k_3) \eta_{\rm end}^6) + 
	5 d k_1^2 (-3 (k_2 + k_3) (k_2^4 - k_2^3 k_3 + k_2^2 k_3^2 - k_2 k_3^3 + \nonumber \\
	&&
	k_3^4) (3 k_2^4 + 18 k_2^3 k_3 + 38 k_2^2 k_3^2 + 18 k_2 k_3^3 + 
	3 k_3^4) \eta_{\rm end}^2 - 
	k_2^2 k_3^2 (k_2 + k_3)^3 (3 k_2^4 + \nonumber \\
	&&9 k_2^3 k_3 - 8 k_2^2 k_3^2 + 
	9 k_2 k_3^3 + 3 k_3^4) \eta_{\rm end}^4 - 
	k_2^4 k_3^4 (k_2 + k_3)^5 \eta_{\rm end}^6 + 
	3 k_1 (k_2 + k_3)^2 \nonumber \\
	&&(9 (k_2 + k_3)^4 - 
	6 (k_2^6 + k_2^5 k_3 - 4 k_2^4 k_3^2 - 2 k_2^3 k_3^3 - 4 k_2^2 k_3^4 +
	k_2 k_3^5 + k_3^6) \eta_{\rm end}^2 -\nonumber \\
	&& 
	k_2^2 k_3^2 (k_2 + k_3)^2 (k_2^2 - 4 k_2 k_3 + k_3^2) \eta_{\rm end}^4) + 
	k_1^5 (27 (k_2^2 + k_3^2) + 
	18 (k_2^4 + k_2^2 k_3^2 +\nonumber \\
	&& k_3^4) \eta_{\rm end}^2 + 
	9 k_2^2 k_3^2 (k_2^2 + k_3^2) \eta_{\rm end}^4 + 
	4 k_2^4 k_3^4 \eta_{\rm end}^6) + 
	3 k_1^4 (k_2 + k_3) (54 (k_2^2 + k_3^2) + \nonumber \\
	&&
	3 (7 k_2^4 + 5 k_2^3 k_3 + 7 k_2^2 k_3^2 + 5 k_2 k_3^3 + 
	7 k_3^4) \eta_{\rm end}^2 + 
	k_2^2 k_3^2 (11 k_2^2 + 7 k_2 k_3 + \nonumber \\
	&&11 k_3^2) \eta_{\rm end}^4 + 
	5 k_2^4 k_3^4 \eta_{\rm end}^6) + 
	2 k_1^2 (k_2 + k_3) (81 (k_2 + k_3)^4 + 
	9 (k_2^6 + 8 k_2^5 k_3 \nonumber \\
	&&+ 17 k_2^4 k_3^2 + 19 k_2^3 k_3^3 + 
	17 k_2^2 k_3^4 + 8 k_2 k_3^5 + k_3^6) \eta_{\rm end}^2 + 
	3 k_2^2 k_3^2 (3 k_2^4 + 15 k_2^3 k_3 \nonumber \\
	&& + 14 k_2^2 k_3^2 + 
	15 k_2 k_3^3 + 3 k_3^4) \eta_{\rm end}^4 + 
	5 k_2^4 k_3^4 (k_2 + k_3)^2 \eta_{\rm end}^6) + 
	2 k_1^3 (45 (3 k_2^4 +\nonumber \\
	&& 9 k_2^3 k_3 + 8 k_2^2 k_3^2 + 9 k_2 k_3^3 + 
	3 k_3^4) + 
	3 (k_2^2 + k_3^2) (12 k_2^4 + 45 k_2^3 k_3 + 56 k_2^2 k_3^2 + \nonumber \\
	&& 
	45 k_2 k_3^3 + 12 k_3^4) \eta_{\rm end}^2 + 
	k_2^2 k_3^2 (21 k_2^4 + 72 k_2^3 k_3 + 82 k_2^2 k_3^2 + 
	72 k_2 k_3^3 + \nonumber \\
	&& 21 k_3^4) \eta_{\rm end}^4 + 
	10 k_2^4 k_3^4 (k_2 + k_3)^2 \eta_{\rm end}^6))), \\
	\mathcal{G}_2(\k_1, \k_2, \k_3) &=& -\frac{H^2}{8\,\epsilon\, k_1\, k_2^5\, k_3^5\, k_T^7\, \eta_{\rm end}^6} (k_1^2 - k_2^2 - 
	k_3^2) (k_2^2 k_3^2 k_T^4 \eta_{\rm end}^2 (-9 (k_1^2 + 3 k_1 (k_2 + k_3) + \nonumber \\
	&& 2 (k_2^2 + 3 k_2 k_3 + k_3^2)) + 
	3 (3 k_1 k_2 k_3 (k_2 + k_3) + 
	k_1^2 (k_2^2 + k_3^2) - (k_2^2 + k_3^2) (k_2^2 + \nonumber \\
	&&3 k_2 k_3 + 
	k_3^2)) \eta_{\rm end}^2 + (3 k_1 k_2^3 (k_1 + k_2)^2 + 
	9 k_1 k_2^3 (k_1 + k_2) k_3 + k_2^2 (k_1 + k_2) \nonumber \\
	&&(8 k_1 + k_2) k_3^2 + 
	3 k_1 (k_1^2 + 3 k_1 k_2 + 3 k_2^2) k_3^3 + (6 k_1^2 + 9 k_1 k_2 + 
	k_2^2) k_3^4 + 3 k_1 k_3^5)\nonumber \\
	&& \eta_{\rm end}^4 + 
	k_2^2 k_3^2 (k_1 + k_2 + 
	k_3) (2 k_1 k_2 (k_1 + k_2) + (k_1 + k_2) (2 k_1 + k_2) k_3 + (2 k_1 + \nonumber \\
	&& 
	k_2) k_3^2) \eta_{\rm end}^6) - 
	5 d (18 k_1^3 (3 k_2^2 (k_1 + k_2)^2 (k_1 + 5 k_2) + 
	21 k_2^2 (k_1 + k_2) (k_1 + 5 k_2) k_3 + \nonumber \\
	&&(k_1 + 4 k_2) (3 k_1^2 + 
	9 k_1 k_2 + 70 k_2^2) k_3^2 + 
	7 (3 k_1^2 + 18 k_1 k_2 + 40 k_2^2) k_3^3 + 
	3 (11 k_1 + \nonumber \\
	&&35 k_2) k_3^4 + 15 k_3^5) + 
	3 (21 k_1^2 (k_2 + k_3)^2 (k_2^2 + k_3^2) (k_2^4 + 2 k_2^3 k_3 + 
	4 k_2^2 k_3^2 + 2 k_2 k_3^3 + \nonumber \\
	&&k_3^4) + 
	2 (k_2 + k_3)^2 (k_2^4 - k_2^3 k_3 + k_2^2 k_3^2 - k_2 k_3^3 + 
	k_3^4) (3 k_2^4 + 18 k_2^3 k_3 + 38 k_2^2 k_3^2 + \nonumber \\
	&&18 k_2 k_3^3 + 
	3 k_3^4) + 3 k_1^6 (7 k_2^4 + 6 k_2^2 k_3^2 + 7 k_3^4) + 
	3 k_1^5 (k_2 + k_3) (25 k_2^4 + 24 k_2^3 k_3 + \nonumber \\
	&&18 k_2^2 k_3^2 + 
	24 k_2 k_3^3 + 25 k_3^4) + 
	2 k_1^3 (k_2 + k_3) (k_2^2 + k_3^2) (27 k_2^4 + 120 k_2^3 k_3 +\nonumber \\
	&& 
	209 k_2^2 k_3^2 + 120 k_2 k_3^3 + 27 k_3^4) + 
	2 k_1^4 (k_2^2 + k_3^2) (48 k_2^4 + 189 k_2^3 k_3 + 
	242 k_2^2 k_3^2 + \nonumber \\
	&&189 k_2 k_3^3 + 48 k_3^4) + 
	k_1 (k_2 + k_3) (15 k_2^8 + 48 k_2^7 k_3 + 34 k_2^6 k_3^2 + 
	92 k_2^5 k_3^3 + 22 k_2^4 k_3^4 + \nonumber \\
	&&92 k_2^3 k_3^5 + 
	34 k_2^2 k_3^6 + 48 k_2 k_3^7 + 15 k_3^8)) \eta_{\rm end}^2 + k_T^2 (9 k_1 k_2^5 (k_1 + k_2)^4 + 
	45 k_1 k_2^5 \nonumber \\
	&&(k_1 + k_2)^3 k_3 + 
	3 k_2^4 (k_1 + k_2)^2 (13 k_1^2 + 24 k_1 k_2 + 2 k_2^2) k_3^2 + 
	15 k_2^4 (k_1 + k_2) \nonumber \\
	&&(13 k_1^2 + 2 k_2^2) k_3^3 + 
	k_2^2 (39 k_1^4 + 195 k_1^3 k_2 + 50 k_1^2 k_2^2 + 64 k_1 k_2^3 + 
	26 k_2^4) k_3^4 + \nonumber \\
	&&(9 k_1^5 + 45 k_1^4 k_2 + 150 k_1^3 k_2^2 + 
	195 k_1^2 k_2^3 + 64 k_1 k_2^4 + 4 k_2^5) k_3^5 + (36 k_1^4 + 
	135 k_1^3 k_2 + \nonumber \\
	&&189 k_1^2 k_2^2 + 30 k_1 k_2^3 + 
	26 k_2^4) k_3^6 + 
	3 (18 k_1^3 + 45 k_1^2 k_2 + 28 k_1 k_2^2 + 10 k_2^3) k_3^7 +\nonumber \\
	&& 
	3 (12 k_1^2 + 15 k_1 k_2 + 2 k_2^2) k_3^8 + 9 k_1 k_3^9) \eta_{\rm end}^4 +
	k_2^2 k_3^2 k_T^4 (2 k_2^2 k_3^2 (k_2 + k_3)^2 + 
	3 k_1^3 \nonumber \\
	&&(k_2^3 + k_3^3) + 
	k_1^2 (6 k_2^4 + 9 k_2^3 k_3 + 22 k_2^2 k_3^2 + 9 k_2 k_3^3 + 
	6 k_3^4) + 
	3 k_1 (k_2^5 + 3 k_2^4 k_3 + \nonumber \\
	&&3 k_2 k_3^4 + k_3^5)) \eta_{\rm end}^6 + 
	k_1 k_2^4 k_3^4 (k_2 + k_3) k_T^6 \eta_{\rm end}^8)),
	\end{eqnarray}
	\begin{eqnarray}
	\mathcal{G}_3(\k_1, \k_2, \k_3) &=&  \frac{9\,H^2}{32\,\epsilon\, k_1^3\, k_2^5\, k_T^2 k_3^5\, \eta_{\rm end}^4}(k_1^4 + k_2^4 + 6 k_2^2 k_3^2 + 
	k_3^4 - 2 k_1^2 (k_2^2 + k_3^2)) ((k_1 + 
	k_2)^2 \nonumber\\
	&&(3 (\gamma_{\rm E}-1) k_1^3 - 3 k_1^2 k_2 - k_2^3) + (k_1 + 
	k_2) ((6 \gamma_{\rm E} - 9) k_1^3 - 6 k_1^2 k_2 - 
	2 k_2^3) k_3 + \nonumber\\
	&&(3 (\gamma_{\rm E}-3) k_1^3 - 6 k_1^2 k_2 - 
	2 k_1 k_2^2 - 2 k_2^3) k_3^2 - 2 (2 k_1^2 + k_1 k_2 + k_2^2) k_3^3 -\nonumber\\
	&& 2 (k_1 + k_2) k_3^4 - k_3^5 + 
	3 k_1^3 k_T^2 \,\log[-k_T \eta_{\rm end}]),\\
	\mathcal{G}_4(\k_1, \k_2, \k_3) &=& \frac{-9\,H^2}{32\,\epsilon\,k1 k_2^5 k_T^2 k_3^5\,\eta_{\rm end}^4} (k_1 - k_2 - k_3) (k_1 + k_2 - k_3) (k_1 - k_2 + k_3) \nonumber\\
	&&(3 (k_1 + k_2)^2 (3 (\gamma_{\rm E}-1) k_1^2 + 
	3 (\gamma_{\rm E}-2) k_1 k_2 - 2 k_2^2) + 
	9 (k_1 + k_2) \nonumber\\
	&&((3 \gamma_{\rm E}-4) k_1^2 + (
	3 \gamma_{\rm E}-7) k_1 k_2 - 
	2 k_2^2) k_3 + (3 (9 \gamma_{\rm E}-17) k_1^2 + \nonumber\\
	&&27 (\gamma_{\rm E}-3) k_1 k_2 - 22 k_2^2) k_3^2 + 
	3 ((3 \gamma_{\rm E}-10) k_1 - 6 k_2) k_3^3 - 6 k_3^4 +\nonumber\\ 
	&&9 k_1 k_T^3\, \log[-k_T \eta_{\rm end}]),\\
	\mathcal{G}_5(\k_1, \k_2, \k_3) &=& -\frac{H^2}{32\,\epsilon\, k_1^3 \,k_2^5\, k_3^5\, k_T^7\, \eta_{\rm end}^6} ((k_1^2 - k_2^2)^2 + 2 (k_1^2 + k_2^2) k_3^2 - 
	3 k_3^4) (k_2^2 k_T^4 \eta_{\rm end}^2 \nonumber \\
	&&(9 (3 (k_1 + k_2)^2 (k_1^2 + k_1 k_2 + k_2^2) + 
	9 (k_1 + k_2) (k_1^2 + k_1 k_2 + k_2^2) k_3 + \nonumber\\
	&& 2 (5 k_1^2 + 6 k_1 k_2 + 5 k_2^2) k_3^2 + 6 (k_1 + k_2) k_3^3 + 
	2 k_3^4) + 
	3 (6 k_1^2 k_2^2 (k_1 + k_2)^2 + \nonumber\\ 
	&&18 k_1^2 k_2^2 (k_1 + k_2) k_3 + (3 k_1^4 + 9 k_1^3 k_2 + 
	26 k_1^2 k_2^2 + 6 k_1 k_2^3 + 2 k_2^4) k_3^2 + \nonumber\\
	&&3 (3 k_1^3 + 6 k_1^2 k_2 + 4 k_1 k_2^2 + 2 k_2^3) k_3^3 + 
	2 (4 k_1^2 + 3 k_1 k_2 + 3 k_2^2) k_3^4 + 3 (k_1 + k_2) k_3^5 + \nonumber\\ 
	&&k_3^6) \eta_{\rm end}^2 + 
	k_3^2 (3 k_1^2 k_2^2 (k_1 + k_2) (2 k_1 + k_2) + 
	9 k_1^2 k_2^2 (2 k_1 + k_2) k_3 + (3 k_1^4 + \nonumber\\
	&&9 k_1^3 k_2 + 
	16 k_1^2 k_2^2 + 3 k_1 k_2^3 + k_2^4) k_3^2 + 
	3 (k_1 + k_2) (2 k_1^2 + k_1 k_2 + k_2^2) k_3^3 +\nonumber\\
	&& (3 k_1^2 + 
	2 k_2^2) k_3^4) \eta_{\rm end}^4 + 
	2 k_1^2 k_2^2 k_3^4 (k_1 + k_3) (k_1 + k_2 + k_3) \eta_{\rm end}^6) - \nonumber\\
	&&5 d (9 k_1^3 (9 k_1^5 + 63 k_1^4 (k_2 + k_3) + 
	210 k_1^2 (k_2 + k_3) (k_2^2 + 3 k_2 k_3 + k_3^2) + \nonumber\\
	&&6 k_1^3 (29 k_2^2 + 63 k_2 k_3 + 29 k_3^2) + 
	35 k_1 (3 k_2^4 + 18 k_2^3 k_3 + 38 k_2^2 k_3^2 + 18 k_2 k_3^3 + 
	3 k_3^4) + \nonumber\\ 
	&&5 (k_2 + k_3) (3 k_2^4 + 18 k_2^3 k_3 + 38 k_2^2 k_3^2 + 
	18 k_2 k_3^3 + 3 k_3^4)) + 
	3 (3 k_2^2 (k_1 + k_2)^4 \nonumber\\
	&&(3 k_1^4 + 9 k_1^3 k_2 - 4 k_1^2 k_2^2 + 
	3 k_1 k_2^3 + k_2^4) + 
	21 k_2^2 (k_1 + k_2)^3 (3 k_1^4 + 9 k_1^3 k_2 - \nonumber\\
	&&4 k_1^2 k_2^2 + 
	3 k_1 k_2^3 + k_2^4) k_3 + (k_1 + k_2)^2 (9 k_1^6 + 45 k_1^5 k_2 + 
	243 k_1^4 k_2^2 + 477 k_1^3 k_2^3 - \nonumber\\
	&&196 k_1^2 k_2^4 + 
	154 k_1 k_2^5 + 56 k_2^6) k_3^2 + 
	7 (k_1 + k_2) (9 k_1^6 + 45 k_1^5 k_2 + 99 k_1^4 k_2^2 + \nonumber\\
	&&45 k_1^3 k_2^3 - 4 k_1^2 k_2^4 + 10 k_1 k_2^5 + 
	8 k_2^6) k_3^3 + (150 k_1^6 + 672 k_1^5 k_2 + 1001 k_1^4 k_2^2 + \nonumber\\
	&&287 k_1^3 k_2^3 + 21 k_1 k_2^5 + 21 k_2^6) k_3^4 + (159 k_1^5 + 
	441 k_1^4 k_2 + 239 k_1^3 k_2^2 + 42 k_1^2 k_2^3 + \nonumber\\
	&&21 k_1 k_2^4 + 
	6 k_2^5) k_3^5 + 
	21 (k_1 + k_2) (4 k_1^3 + 3 k_1^2 k_2 + 5 k_1 k_2^2 + 
	k_2^3) k_3^6 + (45 k_1^3 + \nonumber\\
	&&168 k_1^2 k_2 + 266 k_1 k_2^2 + 
	56 k_2^3) k_3^7 + 14 (3 k_1^2 + 9 k_1 k_2 + 4 k_2^2) k_3^8 + 
	21 (k_1 + k_2) k_3^9 + \nonumber\\
	&&3 k_3^{10}) \eta_{\rm end}^2 + (k_1 + k_2 + 
	k_3)^2 (9 k_1^2 k_2^4 (k_1 + k_2)^4 + 
	45 k_1^2 k_2^4 (k_1 + k_2)^3 k_3 +\nonumber\\ 
	&&3 k_2^2 (k_1 + k_2)^2 (3 k_1^4 + 9 k_1^3 k_2 + 17 k_1^2 k_2^2 + 
	3 k_1 k_2^3 + k_2^4) k_3^2 + 
	15 k_2^2 (k_1 + k_2)\nonumber
	\end{eqnarray}
	\begin{eqnarray}
	&& (3 k_1^4 + 9 k_1^3 k_2 - 7 k_1^2 k_2^2 + 
	3 k_1 k_2^3 + k_2^4) k_3^3 + (9 k_1^6 + 45 k_1^5 k_2 + 
	114 k_1^4 k_2^2 + 30 k_1^3 k_2^3 + \nonumber\\
	&&88 k_1^2 k_2^4 + 5 k_1 k_2^5 + 
	13 k_2^6) k_3^4 + (36 k_1^5 + 135 k_1^4 k_2 + 138 k_1^3 k_2^2 - 
	60 k_1^2 k_2^3 + 5 k_1 k_2^4 + \nonumber\\
	&&2 k_2^5) k_3^5 + (54 k_1^4 + 
	135 k_1^3 k_2 + 72 k_1^2 k_2^2 + 60 k_1 k_2^3 + 13 k_2^4) k_3^6 + 
	3 (12 k_1^3 + 15 k_1^2 k_2 + \nonumber\\
	&&5 k_1 k_2^2 + 5 k_2^3) k_3^7 + 
	3 (3 k_1^2 + k_2^2) k_3^8) \eta_{\rm end}^4 + 
	k_2^2 k_3^2 (k_1 + k_2 + k_3)^4 (3 k_1 k_2^2 k_3^2 (k_2 + k_3) +\nonumber\\
	&& 
	k_2^2 k_3^2 (k_2 + k_3)^2 + 3 k_1^4 (k_2^2 + k_3^2) + 
	3 k_1^3 (k_2 + k_3) (2 k_2^2 + k_2 k_3 + 2 k_3^2) + 
	k_1^2 (3 k_2^4 + \nonumber\\
	&&9 k_2^3 k_3 - 16 k_2^2 k_3^2 + 9 k_2 k_3^3 + 
	3 k_3^4)) \eta_{\rm end}^6 + 
	k_1^2 k_2^4 k_3^4 (k_1 + k_2 + k_3)^6 \eta_{\rm end}^8)),\\
	\mathcal{G}_6(\k_1, \k_2, \k_3) &=& -\frac{H^2}{32\,\epsilon\, k_1^3\, k_2^5\, k_3^5\, k_T^7 \, \eta_{\rm end}^6}((k_1^2 - k_2^2)^2 + 2 (k_1^2 + k_2^2) k_3^2 - 
	3 k_3^4) (k_3^2 (k_1 + k_2 + 
	k_3)^4 \eta_{\rm end}^2 \nonumber\\
	&&(9 (3 k_1^4 + 9 k_1^3 (k_2 + k_3) + 
	2 k_1^2 (5 k_2^2 + 9 k_2 k_3 + 6 k_3^2) + 
	3 k_1 (2 k_2^3 + 4 k_2^2 k_3 + 6 k_2 k_3^2 + \nonumber\\
	&&3 k_3^3) + (k_2 + 
	k_3) (2 k_2^3 + 4 k_2^2 k_3 + 6 k_2 k_3^2 + 3 k_3^3)) + 
	3 (k_2^2 (k_1 + k_2) (3 k_1^3 + 6 k_1^2 k_2 + \nonumber\\
	&&2 k_1 k_2^2 + k_2^3) + 
	3 k_2^2 (3 k_1^3 + 6 k_1^2 k_2 + 2 k_1 k_2^2 + k_2^3) k_3 + 
	2 (3 k_1^4 + 9 k_1^3 k_2 + 13 k_1^2 k_2^2 +\nonumber\\
	&& 6 k_1 k_2^3 + 
	3 k_2^4) k_3^2 + 
	6 (2 k_1^3 + 3 k_1^2 k_2 + k_1 k_2^2 + k_2^3) k_3^3 + 
	2 (3 k_1^2 + k_2^2) k_3^4) \eta_{\rm end}^2 + \nonumber\\
	&&
	k_2^2 (3 k_1^2 k_2^2 (k_1 + k_2)^2 + 9 k_1^2 k_2^2 (k_1 + k_2) k_3 + 
	2 (k_1 + k_2) (3 k_1^3 + 6 k_1^2 k_2 + 2 k_1 k_2^2 + \nonumber\\
	&&k_2^3) k_3^2 + 
	3 (k_1 + k_2) (3 k_1^2 + k_2^2) k_3^3 + (3 k_1^2 + 
	k_2^2) k_3^4) \eta_{\rm end}^4 + 
	2 k_1^2 k_2^4 (k_1 + k_2) k_3^2\nonumber\\
	&& (k_1 + k_2 + k_3) \eta_{\rm end}^6) - 
	5 d (9 k_1^3 (9 k_1^5 + 63 k_1^4 (k_2 + k_3) + 
	210 k_1^2 (k_2 + k_3) (k_2^2 + \nonumber\\
	&&3 k_2 k_3 + k_3^2) + 
	6 k_1^3 (29 k_2^2 + 63 k_2 k_3 + 29 k_3^2) + 
	35 k_1 (3 k_2^4 + 18 k_2^3 k_3 + 38 k_2^2 k_3^2 + \nonumber\\
	&&18 k_2 k_3^3 + 
	3 k_3^4) + 
	5 (k_2 + k_3) (3 k_2^4 + 18 k_2^3 k_3 + 38 k_2^2 k_3^2 + 
	18 k_2 k_3^3 + 3 k_3^4)) + \nonumber\\
	&&
	3 (3 k_2^2 (k_1 + k_2)^4 (3 k_1^4 + 9 k_1^3 k_2 - 4 k_1^2 k_2^2 + 
	3 k_1 k_2^3 + k_2^4) + 
	21 k_2^2 (k_1 + k_2)^3 \nonumber\\
	&&(3 k_1^4 + 9 k_1^3 k_2 - 4 k_1^2 k_2^2 + 
	3 k_1 k_2^3 + k_2^4) k_3 + (k_1 + k_2)^2 (9 k_1^6 + 45 k_1^5 k_2 + 
	243 k_1^4 k_2^2 + \nonumber\\
	&&477 k_1^3 k_2^3 - 196 k_1^2 k_2^4 + 
	154 k_1 k_2^5 + 56 k_2^6) k_3^2 + 
	7 (k_1 + k_2) (9 k_1^6 + 45 k_1^5 k_2 + \nonumber\\
	&&99 k_1^4 k_2^2 + 
	45 k_1^3 k_2^3 - 4 k_1^2 k_2^4 + 10 k_1 k_2^5 + 
	8 k_2^6) k_3^3 + (150 k_1^6 + 672 k_1^5 k_2 + \nonumber\\
	&&1001 k_1^4 k_2^2 + 
	287 k_1^3 k_2^3 + 21 k_1 k_2^5 + 21 k_2^6) k_3^4 + (159 k_1^5 + 
	441 k_1^4 k_2 + 239 k_1^3 k_2^2 + \nonumber\\
	&&42 k_1^2 k_2^3 + 21 k_1 k_2^4 + 
	6 k_2^5) k_3^5 + 
	21 (k_1 + k_2) (4 k_1^3 + 3 k_1^2 k_2 + 5 k_1 k_2^2 + 
	k_2^3) k_3^6 +\nonumber\\
	&& (45 k_1^3 + 168 k_1^2 k_2 + 266 k_1 k_2^2 + 
	56 k_2^3) k_3^7 + 14 (3 k_1^2 + 9 k_1 k_2 + 4 k_2^2) k_3^8 + \nonumber\\
	&&
	21 (k_1 + k_2) k_3^9 + 3 k_3^{10}) \eta_{\rm end}^2 + (k_1 + k_2 + 
	k_3)^2 (9 k_1^2 k_2^4 (k_1 + k_2)^4 + \nonumber\\
	&&
	45 k_1^2 k_2^4 (k_1 + k_2)^3 k_3 + 
	3 k_2^2 (k_1 + k_2)^2 (3 k_1^4 + 9 k_1^3 k_2 + 17 k_1^2 k_2^2 + 
	3 k_1 k_2^3 + \nonumber\\
	&&k_2^4) k_3^2 + 
	15 k_2^2 (k_1 + k_2) (3 k_1^4 + 9 k_1^3 k_2 - 7 k_1^2 k_2^2 + 
	3 k_1 k_2^3 + k_2^4) k_3^3 + (9 k_1^6 + \nonumber\\
	&&45 k_1^5 k_2 + 
	114 k_1^4 k_2^2 + 30 k_1^3 k_2^3 + 88 k_1^2 k_2^4 + 5 k_1 k_2^5 + 
	13 k_2^6) k_3^4 + (36 k_1^5 + 135 k_1^4 k_2 + \nonumber\\
	&&138 k_1^3 k_2^2 - 
	60 k_1^2 k_2^3 + 5 k_1 k_2^4 + 2 k_2^5) k_3^5 + (54 k_1^4 + 
	135 k_1^3 k_2 + 72 k_1^2 k_2^2 + 60 k_1 k_2^3 + \nonumber\\
	&&13 k_2^4) k_3^6 + 
	3 (12 k_1^3 + 15 k_1^2 k_2 + 5 k_1 k_2^2 + 5 k_2^3) k_3^7 + 
	3 (3 k_1^2 + k_2^2) k_3^8) \eta_{\rm end}^4 + \nonumber\\
	&&
	k_2^2 k_3^2 (k_1 + k_2 + k_3)^4 (3 k_1 k_2^2 k_3^2 (k_2 + k_3) + 
	k_2^2 k_3^2 (k_2 + k_3)^2 + 3 k_1^4 (k_2^2 + k_3^2) +\nonumber\\
	&& 
	3 k_1^3 (k_2 + k_3) (2 k_2^2 + k_2 k_3 + 2 k_3^2) + 
	k_1^2 (3 k_2^4 + 9 k_2^3 k_3 - 16 k_2^2 k_3^2 + 9 k_2 k_3^3 + \nonumber\\
	&&
	3 k_3^4)) \eta_{\rm end}^6 + 
	k_1^2 k_2^4 k_3^4 (k_1 + k_2 + k_3)^6 \eta_{\rm end}^8)).
\end{eqnarray}
$\gamma_{\rm E}$ is the Euler-Mascheroni constant and $k_T \equiv k_1 + k_2 + k_3.$


\providecommand{\href}[2]{#2}\begingroup\raggedright\endgroup
\end{document}